\documentclass[aps,prev,onecolumn,preprintnumbers,floatfix,nofootinbib]{revtex4-1}
\pdfoutput=1
\usepackage{geometry,amsmath,amsfonts}
\usepackage{slashed}
\usepackage{epsfig}
\usepackage{latexsym}
\usepackage{graphicx}
\usepackage{hyperref}
\usepackage[justification=RaggedRight]{caption}
\usepackage{amssymb}
\usepackage{subfig}
\usepackage{color}
\usepackage{multirow}
\usepackage{color}
\usepackage{rotating}
\usepackage{ifthen}
\usepackage{epsfig}

\usepackage[utf8]{inputenc} 
\usepackage[T1]{fontenc}

\begin{document}

\title{WIMPs Below the Radar: Blind Spots and Benchmarks Beyond the Neutrino Floor}
\author{Giorgio Arcadi$^{a,b}$}
\email{giorgio.arcadi@unime.it}
\author{Stefano Profumo$^{c,d}$}
\email{profumo@ucsc.edu}

\vspace{0.1cm}
 \affiliation{
${}^a$ 
 Dipartimento di Scienze Matematiche e Informatiche, Scienze Fisiche e Scienze della Terra, Universita degli Studi di Messina, Viale Ferdinando Stagno d'Alcontres 31, I-98166 Messina, Italy
}

\vspace{0.1cm}
 \affiliation{
${}^b$ 
INFN Sezione di Catania, Via Santa Sofia 64, I-95123 Catania, Italy
}


\vspace{0.1cm}
\affiliation{${}^c$ Department of Physics, University of California, Santa Cruz, 
1156 High St, Santa Cruz, CA 95060, United States of America}
\affiliation{${}^d$ Santa Cruz Institute for Particle Physics, University of California, Santa Cruz, 
1156 High St, Santa Cruz, CA 95060, United States of America}

\begin{abstract}
We investigate benchmark scenarios for Weakly Interacting Massive Particles (WIMPs) that naturally evade current direct detection constraints by featuring suppressed spin-independent cross-sections. Focusing on three representative models—the Singlet-Doublet fermion model, its extension to a Two-Higgs-Doublet plus pseudoscalar sector (2HDM$+a$), and a dark $SU(3)$ gauge model—we systematically analyze the interplay between thermal freeze-out, direct detection blind spots, and radiative corrections. In each case, we identify viable regions of parameter space where the predicted dark matter relic abundance is consistent with observations while elastic scattering rates lie below current exclusion limits and, in some cases, but now always, below the neutrino floor. Loop-induced effects are shown to play a critical role, particularly in scenarios with suppressed tree-level interactions. Our findings demonstrate that models with rich electroweak and scalar sectors can populate the experimentally challenging, yet phenomenologically motivated parameter space between existing constraints and the ultimate sensitivity of current-technology direct detection experiments.
\end{abstract}

\maketitle

\section{Introduction}

The thermal relic density assumption underpins many traditional and simplified WIMP models, as it provides a compelling and predictive framework for explaining the observed abundance of dark matter in the universe. Historically, in the standard picture, WIMPs naturally achieve the correct relic density via the {\em thermal freeze-out} mechanism, whereby dark matter particles were initially in thermal equilibrium with the primordial plasma of the early universe. As the universe expanded and cooled, the annihilation rate of WIMPs became insufficient to balance their production, causing their number density to freeze out at a value uniquely determined by their annihilation cross-section. This elegant mechanism directly links the observed relic abundance of dark matter to the weak-scale couplings and masses of WIMPs through the approximate relation $\langle \sigma v \rangle \sim 3 \times 10^{-26} \ \mathrm{cm^3/s}$, where $\langle \sigma v \rangle$ is the thermally averaged annihilation cross-section \cite{refa6, refa18}. This ``WIMP miracle'' makes WIMPs a theoretically attractive candidate for dark matter, with well-motivated mass ranges spanning from $\sim 1 \ \mathrm{GeV}$ to $\sim 10 \ \mathrm{TeV}$.

Traditional WIMP models often arise naturally in extensions of the Standard Model, with supersymmetry (SUSY) being one of the most extensively studied and well-motivated scenarios. In supersymmetric models, the lightest neutralino—a linear combination of the superpartners of the neutral components of the Higgs and electroweak gauge bosons—provides a natural WIMP candidate with inherent stability guaranteed by R-parity conservation \cite{refa1, refa6, refa16}. Several cosmological and collider constraints systematically refine the viable parameter space of neutralino dark matter, with experimental efforts focusing on probing both spin-independent and spin-dependent interactions below the neutrino floor \cite{refa1, refa3}. Neutralino models are tightly constrained by the thermal relic density requirement, as their annihilation cross-sections are directly influenced by their SU(2) gauge properties, masses, and electroweak couplings, including resonant enhancement effects that occur near the masses of the heavy Higgs bosons or the Z boson \cite{refa2, refa6}.

Simplified WIMP models extend and generalize the concept of thermal freeze-out by focusing on effective or minimal couplings between dark matter particles and Standard Model fields, deliberately eschewing full model dependence and UV completion details. These models retain the fundamental ability to precisely match the observed relic density by parameterizing the interactions through specific mediator particles such as scalar, pseudoscalar, vector, or axial-vector bosons \cite{refa8, refa16}. For instance, in scalar- and vector-mediated models, a single mediator particle facilitates WIMP annihilation into Standard Model particles, maintaining theoretical simplicity while capturing the essential phenomenological features \cite{refa17}. These streamlined frameworks enable systematic and comprehensive exploration of WIMP parameter space, isolating thermal relic constraints in contexts that do not necessarily rely on supersymmetric models or other complex high-energy completions \cite{refa12, refa16}. Additionally, simplified WIMP models are particularly well-suited for ongoing experimental programs examining sub-GeV dark matter, where modifying the velocity- or momentum-dependence of annihilation cross-sections can produce the correct relic abundance while simultaneously evading stringent detection constraints at higher masses \cite{refa8}.

Beyond minimally coupled WIMPs, hybrid or more complex theoretical frameworks add rich phenomenological structure to the connection between thermal relic cross-sections and direct detection signals. For instance, multi-component dark matter models or scenarios involving co-annihilation processes can significantly adjust the annihilation dynamics to still yield the required relic density while producing distinctive and potentially observable experimental signatures \cite{refa12, refa17}. Other compelling scenarios involve inelastic scattering mechanisms, where WIMPs interact with target nuclei only by transitioning to a slightly heavier or lighter mass eigenstate, fundamentally altering the energy recoil spectrum and making detection considerably more challenging using standard experimental setups \cite{refa8}.

Regardless of the specific model implementation or theoretical construction, all these WIMP scenarios must ultimately contend with the formidable challenge posed by the neutrino floor as they approach the sensitivity thresholds of current and next-generation direct detection experiments. These models generically predict that the thermally motivated annihilation cross-section overlaps with the irreducible neutrino-induced backgrounds for WIMPs in conventional mass ranges \cite{refa3, refa9, refa6}
$$\sim 10 \ \mathrm{GeV}\,\,\, \mbox{to} \sim 1 \ \mathrm{TeV}.$$
However, as technological advances progressively bring experiments closer to breaking through this fundamental barrier through innovative strategies like directional detection techniques, sub-GeV detector energy thresholds, and sophisticated statistical separations of signal and background event topologies, thermal relic WIMPs remain one of the primary theoretical motivations driving future experimental designs and detector concepts \cite{refa5, refa7, refa14}.

On the other hand, it is well established that most conventional WIMP models enforce an overly restrictive correlation between the interactions responsible for dark matter annihilation processes in the early universe and elastic scattering cross-sections with nucleons in terrestrial detectors. Consequently, present experimental campaigns have already demonstrated a remarkably high capability of testing and systematically ruling out a broad variety of theoretically motivated models (see e.g. \cite{Arcadi:2017kky,Arcadi:2019lka,Arcadi:2024ukq} for comprehensive reviews). There are nevertheless several promising theoretical possibilities to overcome the aforementioned tension and constraint. First of all, one can consider the so-called ``secluded regime'', consisting of scenarios in which the dark matter can efficiently annihilate into light mediator particles and/or, in theories with rich dark sector phenomenology, light beyond-the-Standard-Model unstable particles that subsequently decay to visible states. In such cases it becomes possible to achieve the requisite efficient dark matter annihilation processes while maintaining the dark matter-nucleon scattering cross-section well below current experimental sensitivity limits (for concrete examples and detailed implementations see e.g. \cite{Arcadi:2016qoz}).

An alternative and particularly intriguing scenario is represented by a natural suppression of the effective coupling responsible for direct detection between dark matter and nucleons in the presence of so-called blind spots. These typically correspond to very specific and often fine-tuned assignments of the fundamental model parameters, which are frequently not stable under quantum radiative corrections. The inclusion of quantum effects emerging at one or more loop levels is therefore crucial to rigorously assess the true capability of direct detection experiments to test and constrain this class of models. A more theoretically robust and interesting case is represented by models in which the interactions between dark matter and nucleons are naturally suppressed because the underlying interaction Lagrangian does not generate, at tree level, effective operators that account for dark matter-nucleon scattering with rates anywhere near present and future experimental sensitivities in the non-relativistic limit. Again, this appealing theoretical picture might be fundamentally altered once effective operators arising at the quantum loop level are properly accounted for in the analysis.

In the present study we focus systematically on three distinct classes of theoretically motivated models that exhibit the remarkable feature that their spin-independent direct-detection cross section—quantifying the coherent elastic interaction between the dark matter candidate and target nucleons—is naturally suppressed by the occurrence of "blind spots" in parameter space. The latter arise for a variety of distinct physical reasons in the three different model frameworks, leading to a vast and viable parameter space that fills the experimentally challenging gap between current sensitivity limits and the fundamental neutrino floor. The existence and detailed theoretical realization of various models populating this visually narrow, but experimentally both interesting and critically important region, constitutes the main result and primary contribution of the present study. The principal take-away message is that the scientific scope and experimental program of future direct detection experiments extending current capabilities into the neutrino floor regime is not only technologically feasible but also theoretically well-motivated and phenomenologically rich.

The remainder of this study is organized as follows: the next three sections present comprehensive analyses of the three model frameworks under detailed scrutiny: The Singlet-Doublet Model (sec.~\ref{sec:mod1}), the Two-Higgs-Doublets-Plus-Pseudoscalar Model (sec.~\ref{sec:mod2}), and the Dark SU(3) model (sec.~\ref{sec:mod3}). The final section \ref{sec:conclusions} presents our detailed discussions, theoretical implications, and conclusions.

\section{Singlet-Doublet Model}\label{sec:mod1}

The so-called singlet-doublet model serves as an excellent benchmark for illustrating the impact of blind spots in direct detection (we refer also to \cite{Bhattiprolu:2025beq} for a very recent study). In its simplest incarnation~\cite{Calibbi:2015nha}, the model extends the SM spectrum by introducing three Weyl fermions: one $SU(2)$ singlet, $S$, and two $SU(2)$ doublets, $D_{L,R}$, with opposite hypercharges $Y = \pm 1/2$. These new fields couple to the Higgs doublet $H$ according to the following Lagrangian:
\begin{equation}
\mathcal{L} = -\frac{1}{2} m_S S^2 - m_D D_L D_R - y_1 D_L H S - y_2 D_R \widetilde{H} S + \mbox{H.c.},
\end{equation}
where $\widetilde{H} = i \sigma_2 H^*$. The doublets are decomposed as follows:
\begin{equation}
D_L = \left(
\begin{array}{c}
N_L \\
E_L
\end{array}
\right), \qquad
D_R = \left(
\begin{array}{c}
- E_R \\
N_R
\end{array}
\right),
\end{equation}
which allows us to define a $3 \times 3$ mass matrix for the electrically neutral new fermions:
\begin{equation}
M = \left(
\begin{array}{ccc}
m_S & \frac{y_1 v_h}{\sqrt{2}} & \frac{y_2 v_h}{\sqrt{2}} \\
\frac{y_1 v_h}{\sqrt{2}} & 0 & m_D \\
\frac{y_2 v_h}{\sqrt{2}} & m_D & 0
\end{array}
\right),
\end{equation}
whose diagonalization yields three Majorana physical states:
\begin{equation}
\chi_i^0 = S U_{i1} + D_L U_{i2} + D_R U_{i3}, \quad i=1,2,3,
\end{equation}
where $U_{ij}$ are the entries of a unitary mixing matrix.

The DM candidate is identified with the lightest of these mass eigenstates, which can achieve the correct relic density through annihilations into SM states mediated by $Z/h$ bosons as well as via gauge interactions. For a detailed discussion of these processes, see e.g.~\cite{Arcadi:2017kky,Arcadi:2019lka,Arcadi:2024ukq}. Focusing on direct detection, the DM exhibits both spin-independent (SI) and spin-dependent (SD) interactions, mediated respectively by the SM Higgs and $Z$ bosons. The relevant cross sections are given by (for definiteness we focus on the case of scattering over protons):
\begin{align}
    & \sigma_{\chi_1^0 p}^{\rm SI} = \frac{\mu_{\chi_1^0 p}^2}{\pi} \frac{m_p^2}{v_h^2} \frac{|y_{h \chi_1^0 \chi_1^0}|^2}{m_h^4} \left| \sum_q f_q \right|^2, \nonumber\\
    & \sigma_{\chi_1^0 p}^{\rm SD} = \frac{3 \mu_{\chi_1^0 p}^2}{\pi m_Z^4} \left| y^A_{Z \chi_1^0 \chi_1^0} \right|^2 \left[ A_u^Z \Delta_u^p + A_d^Z (\Delta_d^p + \Delta_s^p) \right]^2,
\end{align}
with $\mu_{\chi_1^0 p}$ being the DM-proton reduced mass while $f_{q=u,d,s,c,b,t}$ and $\Delta_{q=u,d,s}^p$ are nucleon structure functions associated with SI and SD interactions, respectively. For the former, $f_{q=c,b,t} = 2/27\, f_{TG}$ where the latter is defined by:
\begin{equation}
    \langle N | G_{\mu \nu}G^{\mu \nu}|N \rangle=-m_N \frac{8\pi}{9\alpha_S}f_{TG},\,\,\,\,f_{TG}=1-\sum_{q=u,d,s}f_q
\end{equation}

The couplings of DM to the $h$ and $Z$ bosons can be written analytically as~\cite{Calibbi:2015nha}:
\begin{align}
     & y_{h \chi_1^0 \chi_1^0} = -\frac{(\sin 2\theta\, m_D + m_{\chi_1^0}) y^2 v_h}{m_D^2 + \frac{v_h^2}{2} y^2 + 2 m_S m_{\chi_1^0} - 3 m_{\chi_1^0}^2}, \nonumber \\
     & y_{Z \chi_1^0 \chi_1^0} = -\frac{m_Z v_h y^2 \cos 2\theta (m_{\chi_1^0}^2 - m_D^2)}{2 (m_{\chi_1^0}^2 - m_D^2)^2 + v_h^2 \left[ 2 y^2 \sin 2\theta\, m_{\chi_1^0} m_D + y^2 (m_{\chi_1^0}^2 + m_D^2) \right]}.
\end{align}
From these expressions, it is straightforward to see that the DM couplings can be (though not simultaneously) set to zero by suitable choices of the model parameters. Such parameter choices define the so-called blind spots. We are particularly interested in the case of SI interactions, which exhibit a blind spot when the coupling $y_{h\chi_1^0 \chi_1^0}$ vanishes, i.e.,
\begin{equation}
    m_{\chi_1^0} + m_D \sin 2\theta = 0.
\end{equation}
As previously noted, blind spots typically arise from accidental parameter relations and are not protected by symmetries (in contrast to the "natural" blind spots discussed later). Consequently, the potential impact of loop-induced effects must be assessed. We therefore modify the SI scattering cross-section to include loop corrections:
\begin{align}
\label{eq:SD_full_loop}
& \sigma_{\chi_1^0 p}^{\rm SI} = \frac{\mu_{\chi_1^0 p}^2}{\pi} \frac{m_p^2}{v_h^2} \Bigg| \sum_{q} f_q \left( \frac{g_{h\chi_1^0 \chi_1^0}}{m_h^2} + \frac{\lambda_{hhh}}{m_h^2} C_{\rm triangle,\chi \chi h}^{\rm CP,even} \right) \nonumber\\
& + \sum_{q=u,d,s} f_q C_{1, \rm box}^{\rm CP,even} + \sum_{q=u,d,s,c,b} \frac{3}{4} \left( q(2) + \bar q(2) \right) \left( C_{5, \rm box}^{\rm CP,even} + m_{\chi_1^0} C_{6, \rm box}^{\rm CP,even} \right) + \frac{2}{27} f_{TG} C_{G, \rm box}^{\rm CP,even} \nonumber\\
& + \sum_{q=u,d,s} \frac{m_q}{m_p} f_q C_{q,EW} + \frac{3}{4} \sum_{q=u,d,s,c,b} \left( q(2) + \bar q(2) \right) G_{q,EW} - \frac{8\pi}{9 \alpha_s} f_{TG} C_{G,EW} \Bigg|^2 \, .
\end{align}
The term $C_{\rm triangle}^{\rm CP-even}$ corresponds to contributions from Feynman diagrams with triangle topologies involving the SM Higgs and the BSM electrically neutral fermions. Here, $\lambda_{hhh}$ denotes the SM Higgs trilinear coupling. The terms in the second row of Eq.~\ref{eq:SD_full_loop} represent contributions from box-shaped Feynman diagram topologies. The coefficients $C_{1,5,6,G, \rm box}^{\rm CP-even}$ are adapted from the computation in Ref.~\cite{Ertas:2019dew}, and we use the same notation to facilitate comparison. Additional details are provided in Appendix~\ref{sec:loop_CP_even}. The CP-even label is used to distinguish these results from those discussed below. The last row of Eq.~\ref{eq:SD_full_loop} corresponds to contributions from electroweak (EW) interactions of the DM, determined using the results of~\cite{Belyaev:2022qnf} (see also~\cite{Hisano:2011cs}):
\begin{align}
    & C_{q,EW} = \Bigg\{ \frac{g^2}{64\pi^2 m_W m_h^2} \left| g_{\chi_1^0 \psi W^\pm}^V \right|^2 \Delta_H \left( \frac{m_W^2}{m_{\chi_1^0}^2}, \frac{m_\psi - m_{\chi_1^0}}{m_{\chi_1^0}} \right) \nonumber\\
    & \quad + \frac{g^2}{64\pi^2 m_Z m_h^2} \sum_{j=2,3} \left| g_{\chi_1^0 \chi_j^0 Z}^V \right|^2 \Delta_H \left( \frac{m_Z^2}{m_{\chi_1^0}^2}, \frac{m_{\chi_j^0} - m_{\chi_1^0}}{m_{\chi_1^0}} \right) \Bigg\} \nonumber\\
    & \quad + \frac{g^2}{64\pi^2 m_W^3} \left| g_{\chi_1^0 \psi W^\pm}^V \right|^2 \Delta_S \left( \frac{m_W^2}{m_{\chi_1^0}^2}, \frac{m_\psi - m_{\chi_1^0}}{m_{\chi_1^0}}, V_q^\pm, A_q^\pm \right) \nonumber\\
    & \quad + \frac{g^2}{16\pi^2 m_Z^3} \sum_{j=2,3} \left| g_{\chi_1^0 \chi_j Z}^V \right|^2 \Delta_S \left( \frac{m_Z^2}{m_{\chi_1^0}^2}, \frac{m_{\chi_j^0} - m_{\chi_1^0}}{m_{\chi_1^0}}, V_q^0, A_q^0 \right), \nonumber\\
    & G_{q,EW} = \frac{g^2}{16\pi^2 m_W^3} \left| g_{\chi_1^0 \psi W^\pm}^V \right|^2 \left[ \Delta_{T1} \left( \frac{m_W^2}{m_{\chi_1^0}^2}, \frac{m_\psi - m_{\chi_1^0}}{m_{\chi_1^0}}, \frac{1}{2}, \frac{1}{2} \right) + \Delta_{T2} \left( \frac{m_W^2}{m_{\chi_1^0}^2}, \frac{m_\psi - m_{\chi_1^0}}{m_{\chi_1^0}}, \frac{1}{2}, \frac{1}{2} \right) \right] \nonumber\\
    & \quad + \frac{g^2}{16\pi^2 m_Z^3} \sum_{j=2,3} \left| g_{\chi_1^0 \chi_j^0 W^\pm}^V \right|^2 \left[ \Delta_{T1} \left( \frac{m_Z^2}{m_{\chi_1^0}^2}, \frac{m_{\chi_j^0} - m_{\chi_1^0}}{m_{\chi_1^0}}, V_q^0, A_q^0 \right) + \Delta_{T2} \left( \frac{m_Z^2}{m_{\chi_1^0}^2}, \frac{m_{\chi_j^0} - m_{\chi_1^0}}{m_{\chi_1^0}}, V_q^0, A_q^0 \right) \right].
\end{align}
Here, $\Delta_{H,S,T_1,T_2}$ are loop functions, for which we refer. together with the expressions of the $V^{0,\pm}_q,A_q^{0,\pm}$, to~\cite{Belyaev:2022qnf} for analytical expressions. The function $C_{G,EW}$ is as in~\cite{Hisano:2011cs}, to which we refer for its explicit form.

To analyze the model, we have performed a parameter scan over the following ranges:
\begin{equation}
    m_S \in [1,5000]\,\mbox{GeV}, \quad m_D \in [100,5000]\,\mbox{GeV}, \quad y \in [10^{-3},10], \quad \tan\theta \in [-20,20],
\end{equation}
retaining only those model points consistent with constraints from the invisible widths of the $h$ and $Z$ bosons, electroweak precision tests (EWPT, see e.g.~\cite{Arcadi:2024ukq} for a recent review), and with a DM relic density, computed via the conventional freeze-out paradigm, not exceeding the experimental value. For each viable model point, the DM SI scattering cross-section is computed according to Eq.~\ref{eq:SD_full_loop}.

\begin{figure}
    \centering
    \subfloat{\includegraphics[width=0.5\linewidth]{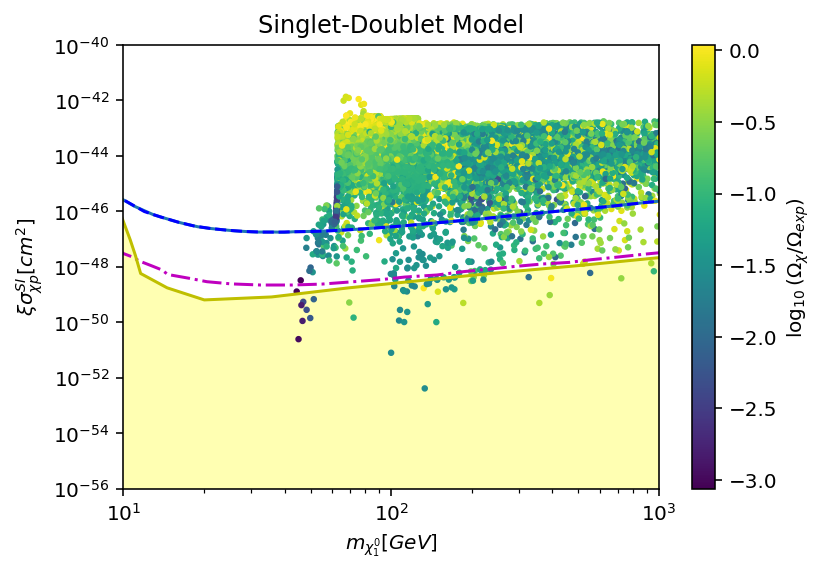}}
    \subfloat{\includegraphics[width=0.5\linewidth]{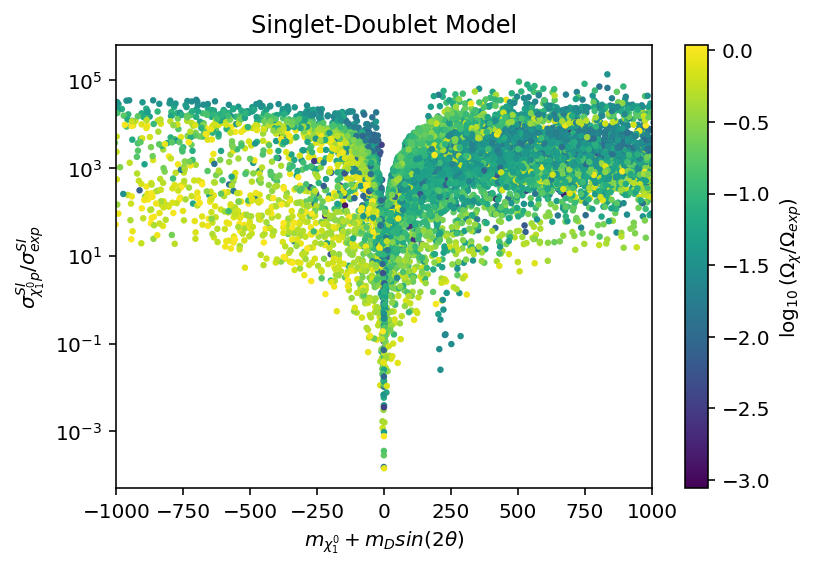}}
    \caption{\footnotesize{{\it Left Panel}: DM scattering cross-section off nucleons as a function of the DM mass, for model assignments (colored points) in the minimal singlet-doublet model. All points correspond to a relic density not exceeding the experimental determination. For models with under-abundant DM, the cross-section is rescaled by $\xi = \Omega_{\chi_1^0}/\Omega_{\rm DM,exp}$. {\it Right Panel}: Same model points shown as a function of $m_{\chi_1^0} + m_D \sin(2\theta)$.}}
    \label{fig:SDscan}
\end{figure}

The results are shown as a function of the DM mass in the left panel of Fig.~\ref{fig:SDscan}. The scaling factor $\xi = \Omega_{\chi_1^0} h^2 / \Omega_{\rm exp} h^2$, with $\Omega_{\rm exp} h^2 = 0.1158$, i.e., the ratio between the DM relic density for a given model point and the experimental value~\cite{Planck:2018vyg}, is applied to account for the possibility that the Majorana fermion constitutes only a fraction of the total DM. The plot displays, as dashed blue and dot-dashed magenta lines, the current exclusion for SI interactions from~\cite{LZ:2022lsv}, as well as the projected exclusion by XLZD \cite{XLZD:2024nsu} \footnote{We have considered, for definiteness the sensitivity corresponding to 200ty of exposure. Ref. \cite{XLZD:2024nsu} provides as well projected exclusion for 1000ty exposure.}, representative of next-generation detectors. The yellow region denotes the neutrino floor \footnote{As discussed e.g. in \cite{OHare:2021utq}, the neutrino floor depends on the detector material. We considered, for our study, the neutrino floor associated to a xenon based detector.}. Model points are color-coded according to the fraction of DM relic density. Notably, no points with DM masses below $50\div 60$\,GeV appear, primarily due to combined bounds from invisible decay widths of the $h$ and $Z$ bosons. Blind spots do not alleviate these limits, as it is not possible to simultaneously cancel the DM couplings to both the Higgs and $Z$ bosons. While most remaining model points are already excluded by current direct detection experiments, some configurations can account for a substantial fraction, or even the totality, of the DM abundance at very small cross-sections. To further characterize these configurations, the right panel of Fig.~\ref{fig:SDscan} displays the ratio of the DM scattering cross-section to the LZ exclusion limit, as a function of $(m_{\chi_1^0} + m_D \sin 2\theta, \sigma_{\chi_1^0\,p}^{\rm SI})$.

As evidenced by the figure, the minimal value of the DM scattering cross-section is reached for $m_{\chi_1^0}+m_D \sin 2\theta=0$, i.e. the blind spot for tree-level interactions. From our numerical study it nevertheless resulted that the inclusion of loop corrections determines a bound on the minimal value of the DM scattering cross-section. For such a reason we see that the majority of the model points lie within the reach of the next generation of facilities and only a small fraction lies within the neutrino floor.


We now extend our analysis to the case of an extended Higgs sector, building on similar approaches in the literature \cite{Berlin:2015wwa,Arcadi:2018pfo}. Starting from the Lagrangian: 
\begin{equation}
\mathcal{L} = -\frac{1}{2}m_S S^2 - m_S D_L D_R - y_1 D_L \Phi_a S - y_2 D_R \widetilde{\Phi}b S + \text{H.c.},
\end{equation}
where $a,b=1,2$, the fermion mass spectrum remains unchanged from the minimal model, leaving the lightest Majorana eigenstate as the DM candidate. In the mass basis, the relevant Lagrangian for DM phenomenology becomes:
\begin{align}
\mathcal{L} &= \overline{\psi^-} \gamma^\mu \left(g^V{W^{\mp}E^{\pm}\chi_i^0} - g^A_{W^{\mp}E^{\pm}\chi_i^0}\gamma_5\right)\chi_i^0 W_\mu^- + \text{H.c.} \nonumber\\
& + \frac{1}{2}\sum_{i,j=1}^3 \overline{\chi_i^0}\gamma^\mu \left(g_{Z \chi_i^0 \chi_j^0}^V - g_{Z \chi_i^0 \chi_j^0}^A \gamma_5\right) \chi_j^0 Z_\mu \nonumber\\
& + \frac{1}{2}\sum_{i,j=1}^{3}\overline{\chi_i^0}\left(y_{h \chi_i^0 \chi_j^0}h + y_{H \chi_i^0 \chi_j^0}H + y_{A \chi_i^0 \chi_j^0}\gamma_5 A\right)\chi_j^0 \nonumber\\
& + \overline{\psi^-} \left(g^S_{H^{\pm}\psi^\mp \chi_i^0} - g^P_{H^{\pm}\psi^\mp \chi_i^0}\gamma_5\right)\chi_i^0 H^- + \text{H.c.} \nonumber\\
& - e A_\mu \overline{\psi^-}\gamma^\mu \psi^- - \frac{g}{2 c_W}(1-2 s^2_W) Z_\mu \overline{\psi^-}\gamma^\mu \psi^- + \text{H.c.},
\end{align}

The spin-independent DM-nucleon scattering cross-section, incorporating radiative corrections, extends the minimal singlet-doublet model case:
\begin{align}
\label{eq:SD_2HDM_full_loop}
\sigma_{\chi_1^0 p}^{\rm SI} &= \frac{\mu_{\chi_1^0 p}^2}{\pi}\frac{m_p^2}{v_h^2}\Bigg| \sum_{q}f_q \left(\frac{y_{h\chi_1^0 \chi_1^0}g_{hqq}}{m_h^2} + \frac{y_{H \chi_1^0 \chi_1^0}g_{Hqq}}{m_H^2}\right) \nonumber\\
& + \sum_q f_q \sum_{\phi=h,H}\frac{1}{m_\phi^2} \left(C_{\rm triangle, \chi \chi \phi}^{\rm CP, even} + C_{\rm triangle,\chi \chi \phi}^{\rm CP, odd}\right) \nonumber\\
& + \sum_{q=u,d,s}f_q \left( C_{1,\rm box}^{\rm CP,even} + C_{1,\rm box}^{\rm CP-odd}\right) \nonumber\\
& + \sum_{q=u,d,s,c,b}\frac{3}{4}\left(q(2)+\bar q (2)\right)\left[C_{5,\rm box}^{\rm CP ,even} + C_{5,\rm box}^{\rm CP ,odd} + m_{\chi_1^0}\left(C_{6,\rm box}^{\rm CP,even} + C_{6,\rm box}^{\rm CP-odd}\right)\right] \nonumber\\
& + \frac{2}{27}f_{TG}\left(C_{G,\rm box}^{\rm CP,even} + C_{G,\rm box}^{\rm CP,odd}\right) \nonumber\\
& + \sum_{q=u,d,s}\frac{m_q}{m_p} f_q C_{q,EW} + \frac{3}{4} \sum_{q=u,d,s,c,b}\left(q(2)+\bar q (2)\right)G_{q,EW} - \frac{8\pi}{9 \alpha_s} f_{TG} C_{G,EW}\Bigg|^2.
\end{align}

The tree-level contribution arises from CP-even Higgs boson couplings to Majorana DM, dependent on parameters $y_{1,2}$ and Higgs mixing elements. Following \cite{Berlin:2015wwa}, we consider four Yukawa coupling configurations (uu/ud/du/dd) with analytical DM couplings (assuming the alignment limit $\beta-\alpha=\pi/2)$:

\textbf{uu configuration:}
\begin{align}
y_{h\chi_1^0 \chi_1^0} &= y^2 v_h \sin^2 \beta \frac{m_{\chi_1^0} + m_D \sin 2\theta}{2 m_D^2 + 4 m_S m_{\chi_1^0} - 6 m_{\chi_1^0}^2 + y^2 v_h^2 \sin^2 \beta}, \nonumber\\
y_{H\chi_1^0 \chi_1^0} &= -\frac{1}{2}y^2 v_h \sin 2\beta \frac{m_{\chi_1^0} + m_D \sin 2\theta}{2 m_D^2 + 4 m_S m_{\chi_1^0} - 6 m_{\chi_1^0}^2 + y^2 v_h^2 \sin^2 \beta}.
\end{align}

\textbf{ud configuration:}
\begin{align}
y_{h \chi_1^0 \chi_1^0} &= y^2 v_h \frac{m_{\chi_1^0} (\sin^2\beta \cos^2\theta + \cos^2\beta \sin^2\theta) + \frac{1}{2}m_D \sin 2\beta \sin 2\theta}{2 m_D^2 + 4 m_S m_{\chi_1^0} - 6 m_{\chi_1^0}^2 + \frac{1}{2}y^2 v_h^2 (1 - \cos 2\beta \cos 2\theta)}, \nonumber\\
y_{H \chi_1^0 \chi_1^0} &= -\frac{1}{2}y^2 v_h \frac{m_{\chi_1^0}\sin 2\beta \cos 2\theta + m_D \cos 2\beta \sin 2\theta}{2 m_D^2 + 4 m_S m_{\chi_1^0} - 6 m_{\chi_1^0}^2 + \frac{1}{2}y^2 v_h^2 (1 - \cos 2\beta \cos 2\theta)}.
\end{align}

\textbf{du configuration:}
\begin{align}
y_{h \chi_1^0 \chi_1^0} &= \frac{1}{2}y^2 v_h \frac{m_{\chi_1^0}(1 + \cos 2\beta \cos 2\theta) + m_D \sin 2\beta \sin 2\theta}{2 m_D^2 + 4 m_S m_{\chi_1^0} - 6 m_{\chi_1^0}^2 + \frac{1}{2}y^2 v_h^2 (1 + \cos 2\beta \cos 2\theta)}, \nonumber\\
y_{H \chi_1^0 \chi_1^0} &= \frac{1}{2}y^2 v_h \frac{m_{\chi_1^0}\sin 2\beta \cos 2\theta - m_D \cos 2\beta \sin 2\theta}{2 m_D^2 + 4 m_S m_{\chi_1^0} - 6 m_{\chi_1^0}^2 + \frac{1}{2}y^2 v_h^2 (1 + \cos 2\beta \cos 2\theta)}.
\end{align}

\textbf{dd configuration:}
\begin{align}
y_{h \chi_1^0 \chi_1^0} &= y^2 v_h \cos^2 \beta \frac{m_{\chi_1^0} + m_D \sin 2\theta}{2 m_D^2 + 4 m_S m_{\chi_1^0} - 6 m_{\chi_1^0}^2 + y^2 v_h^2 \cos^2 \beta}, \nonumber\\
y_{H \chi_1^0 \chi_1^0} &= \frac{1}{2}y^2 v_h \sin 2\beta \frac{m_{\chi_1^0} + m_D \sin \theta}{2 m_D^2 + 4 m_S m_{\chi_1^0} - 6 m_{\chi_1^0}^2 + y^2 v_h^2 \cos^2 \beta}.
\end{align}

We can suppress tree-level contributions through parameter choices ($m_S, m_D, y, \theta, \tan\beta$) that nullify $y_{h \chi_1^0 \chi_1^0}$ or induce destructive interference between Higgs exchange diagrams (for a general study the reader might refer also to \cite{Cabrera:2019gaq}). Loop corrections (particularly from electroweak gauge boson interactions) become significant in blind-spot regions, with contributions represented in Eq.~\ref{eq:SD_2HDM_full_loop}. While complete computation of extended Higgs sector loop contributions remains for future work, our analysis includes known gauge boson-mediated terms analogous to the minimal model.

\begin{figure}[htbp]
\centering
\subfloat[Type-I uu]{\includegraphics[width=0.24\linewidth]{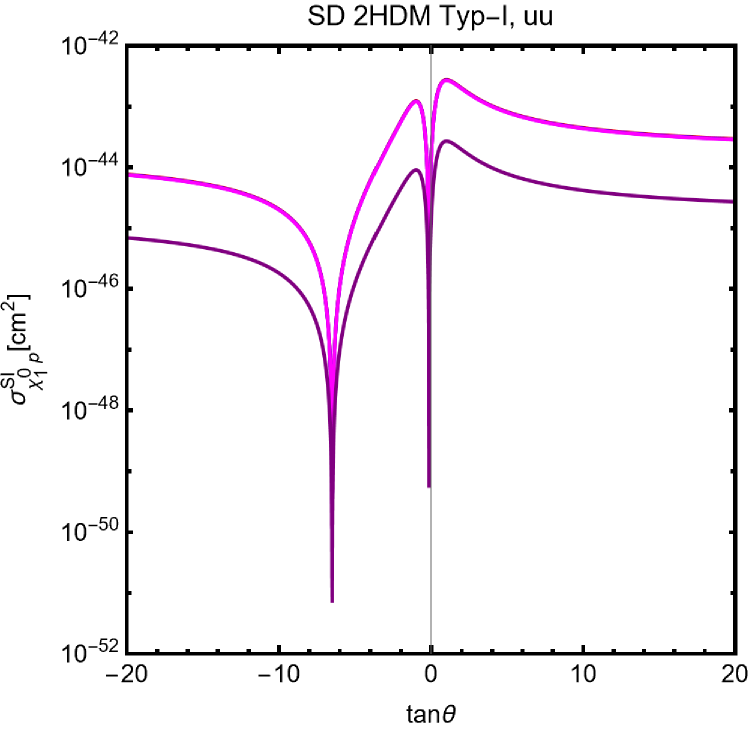}}
\subfloat[Type-I ud]{\includegraphics[width=0.24\linewidth]{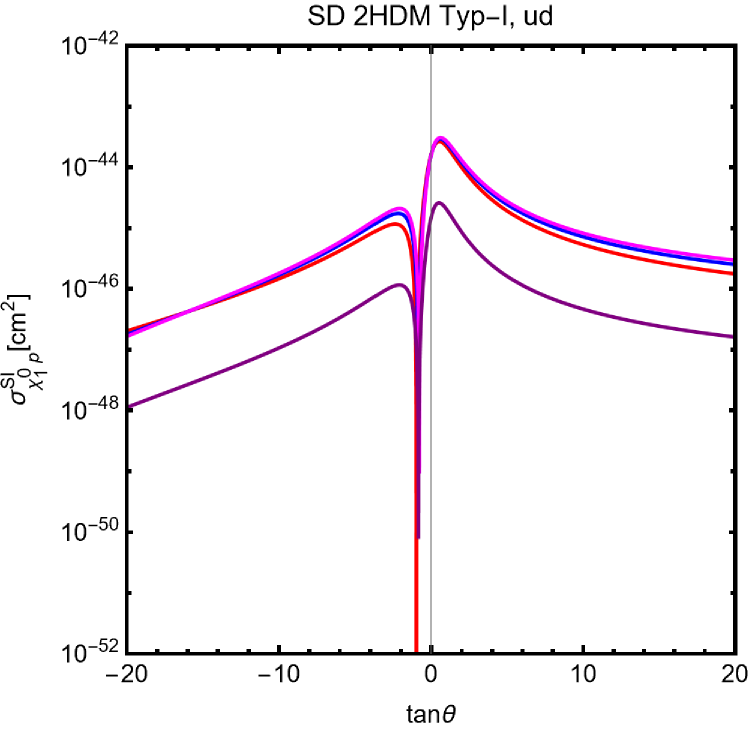}}
\subfloat[Type-I du]{\includegraphics[width=0.24\linewidth]{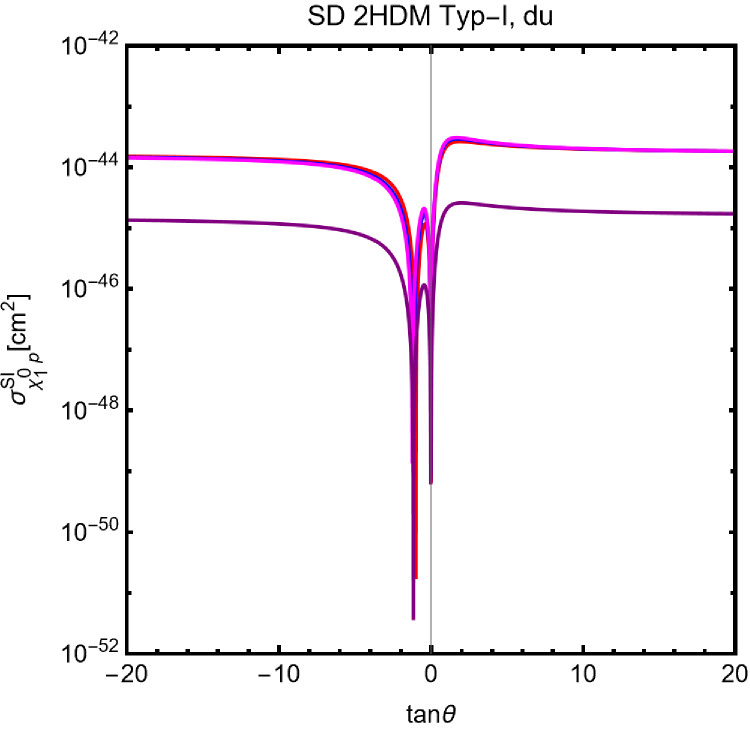}}
\subfloat[Type-I dd]{\includegraphics[width=0.24\linewidth]{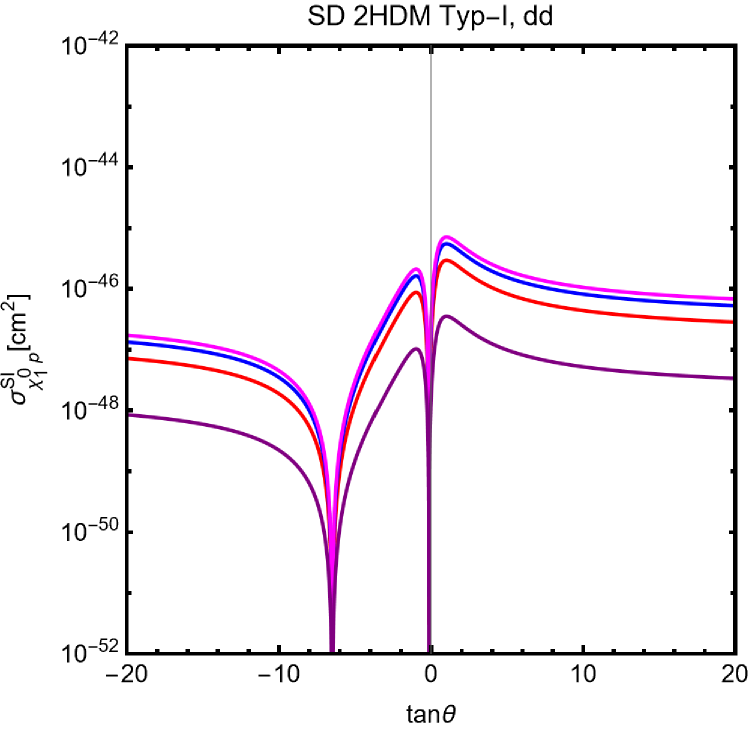}}\
\subfloat[Type-II uu]{\includegraphics[width=0.24\linewidth]{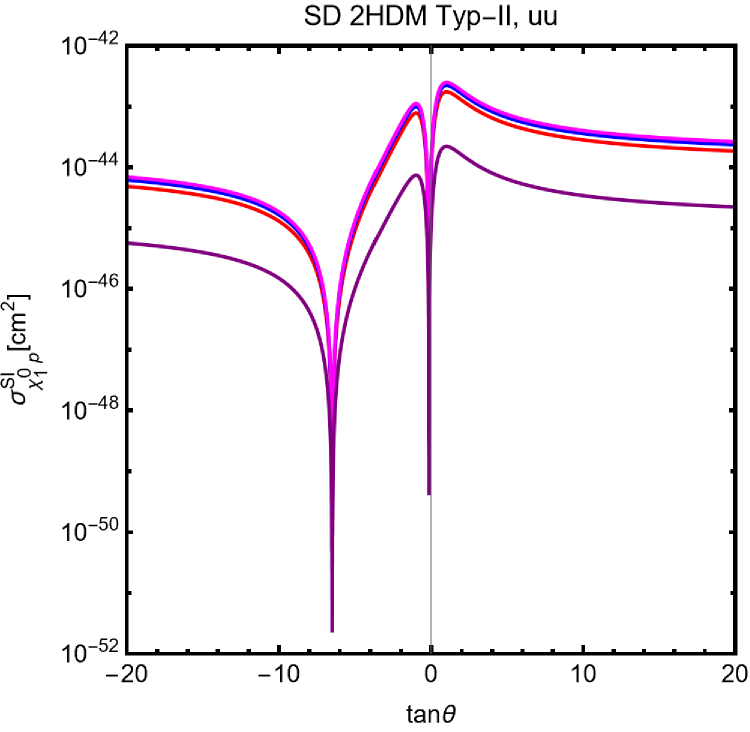}}
\subfloat[Type-II ud]{\includegraphics[width=0.24\linewidth]{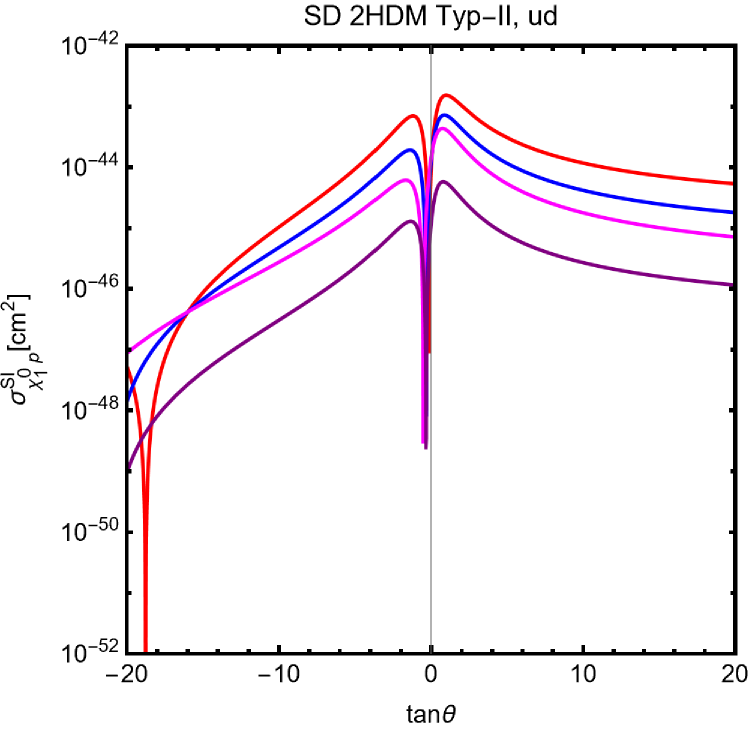}}
\subfloat[Type-II du]{\includegraphics[width=0.24\linewidth]{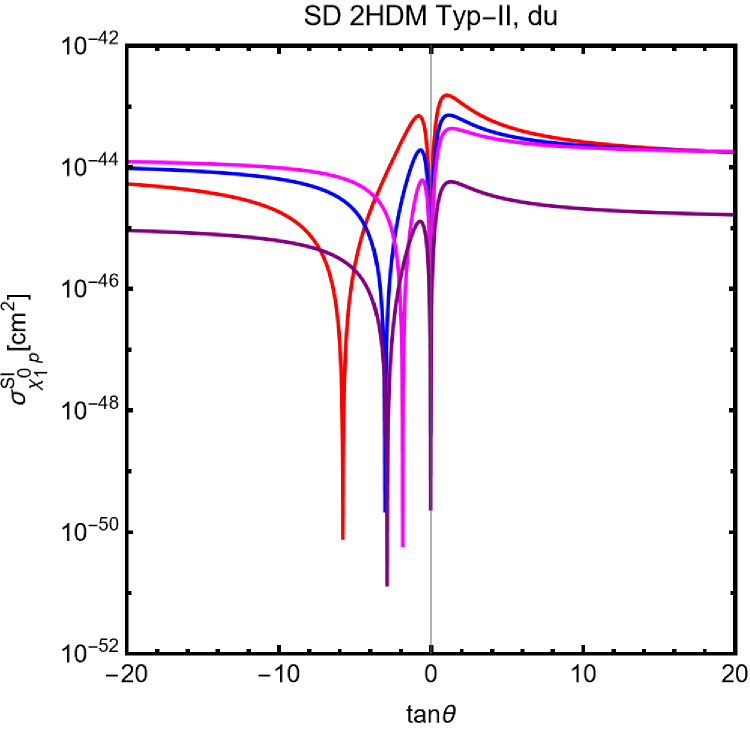}}
\subfloat[Type-II dd]{\includegraphics[width=0.24\linewidth]{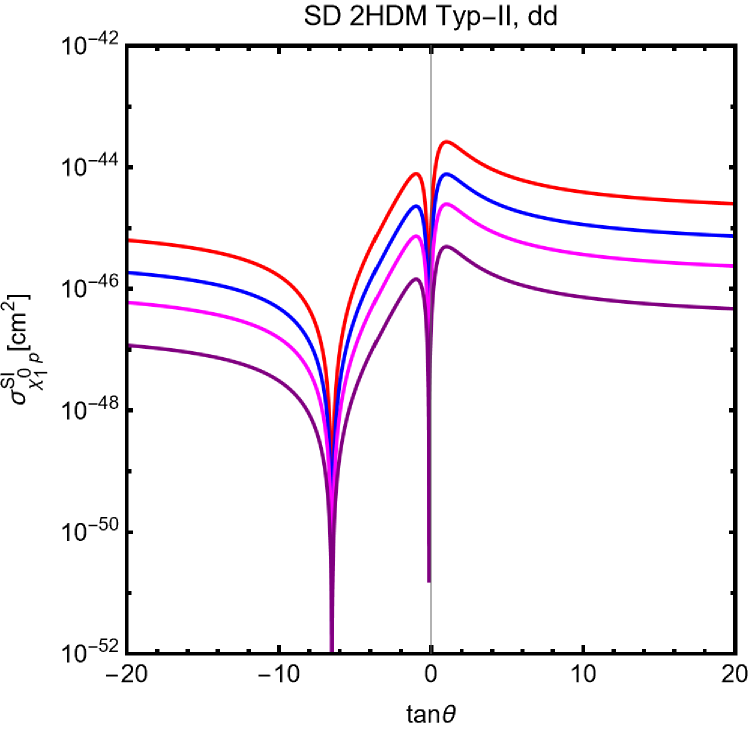}}
\caption{Proton scattering cross-section versus $\tan\theta$ in the Singlet-Doublet+2HDM framework. Parameters: $(m_S,m_D) = (150,500),\text{GeV}$, $\tan\beta=5$. Line colors correspond to $(y,m_H=m_A=m_{H^\pm})$ combinations: (1,200) GeV (red), (1,300) GeV (blue), (1,500) GeV (magenta), (0.5,500) GeV (purple). LZ exclusion limits shown for comparison.}
\label{fig:SD2HDMblind}
\end{figure}

\begin{figure}[htbp]
\centering
\subfloat[Type-I uu]{\includegraphics[width=0.24\linewidth]{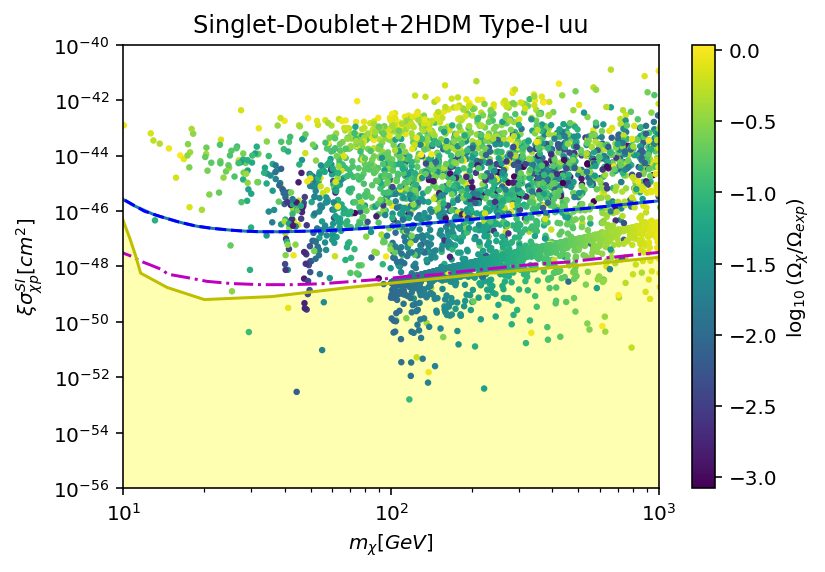}}
\subfloat[Type-I ud]{\includegraphics[width=0.24\linewidth]{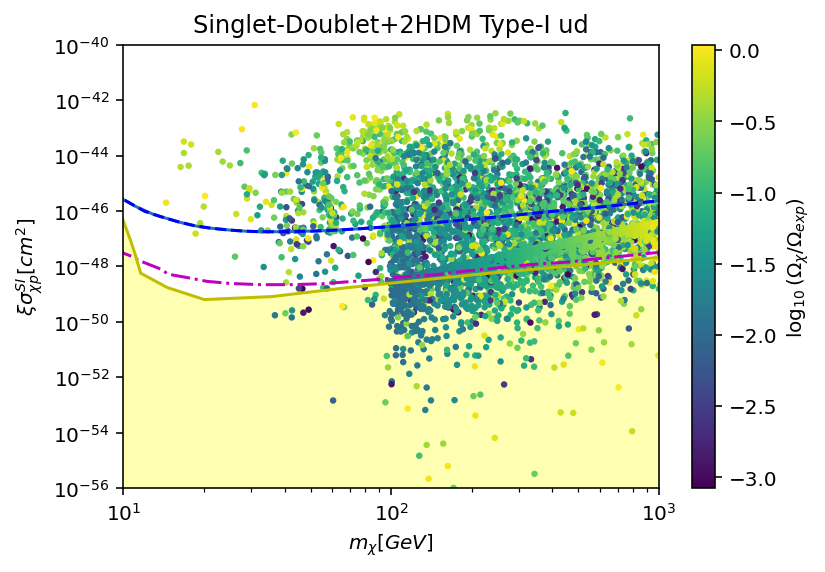}}
\subfloat[Type-I du]{\includegraphics[width=0.24\linewidth]{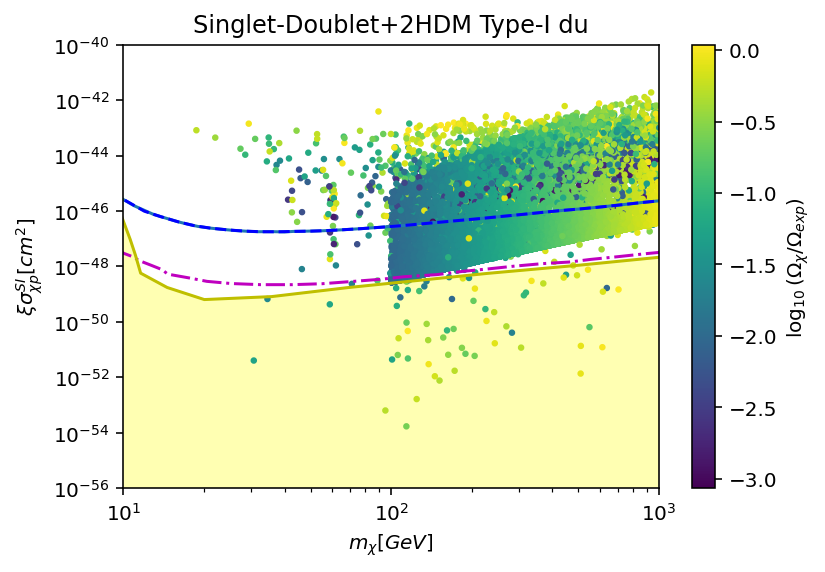}}
\subfloat[Type-I dd]{\includegraphics[width=0.24\linewidth]{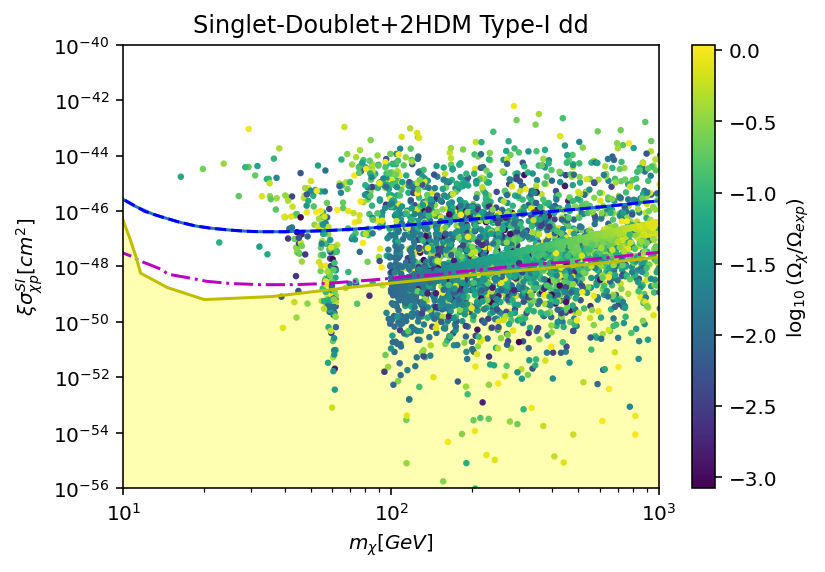}}\
\subfloat[Type-II uu]{\includegraphics[width=0.24\linewidth]{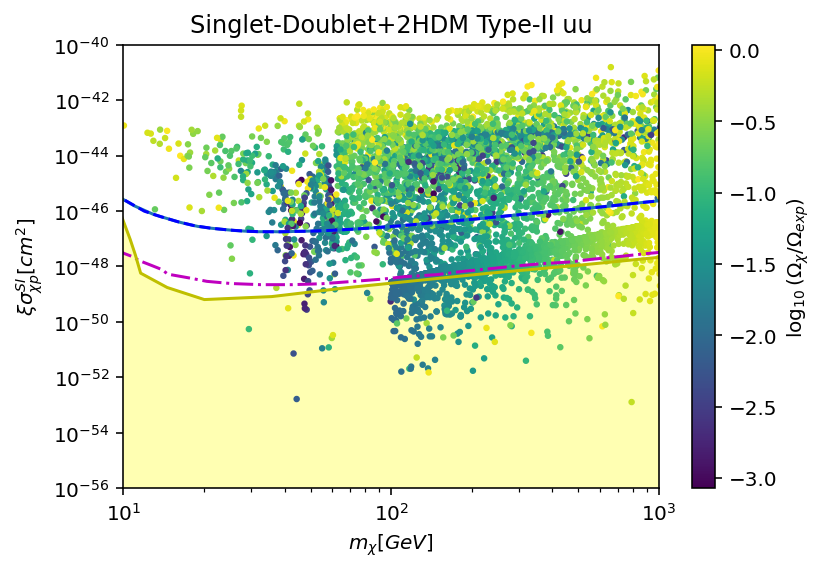}}
\subfloat[Type-II ud]{\includegraphics[width=0.24\linewidth]{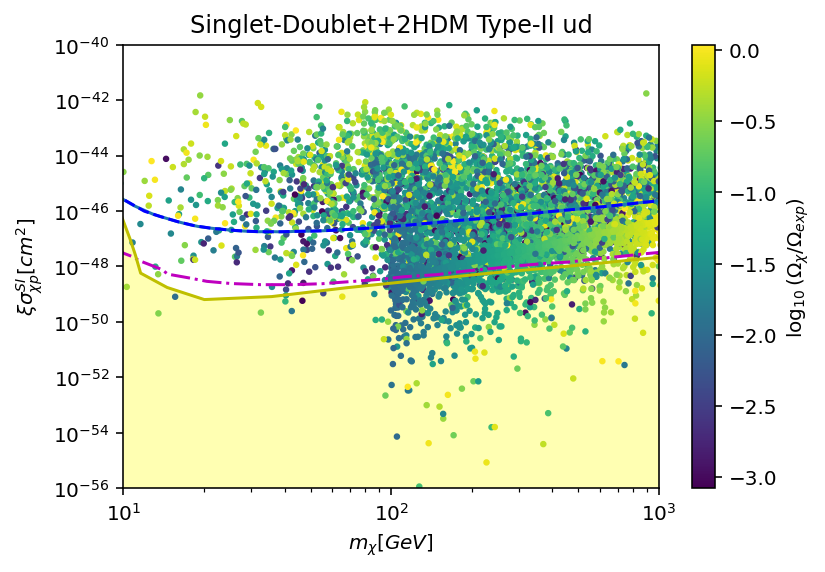}}
\subfloat[Type-II du]{\includegraphics[width=0.24\linewidth]{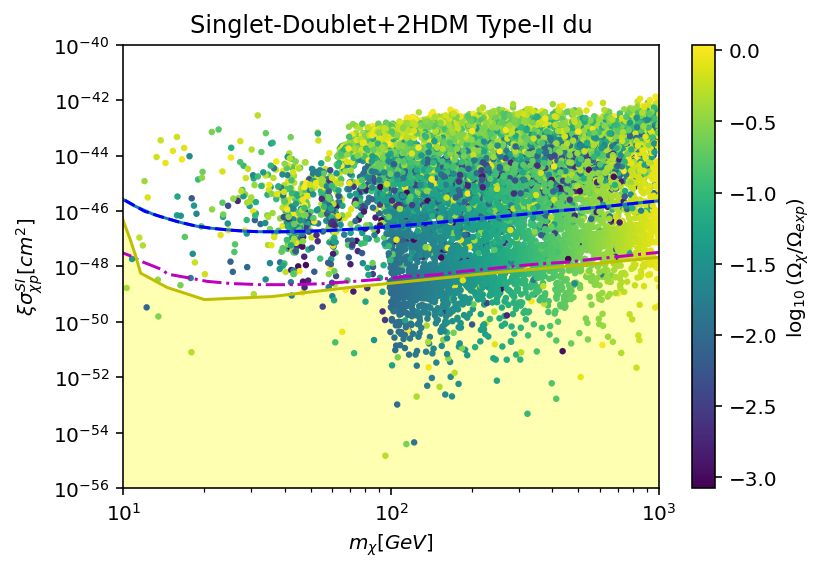}}
\subfloat[Type-II dd]{\includegraphics[width=0.24\linewidth]{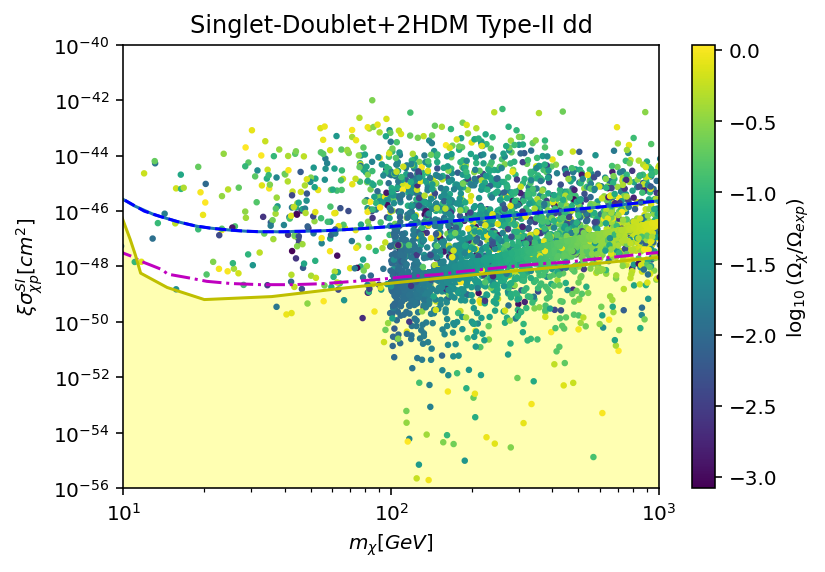}}
\caption{Parameter scan results in the $(m_{\chi_1^0},\xi\sigma_{\chi p}^{\rm SI})$ plane. Color scale indicates relic density $\Omega_{\chi}h^2$ (all points satisfy $\Omega_{\chi} \leq \Omega_{\rm DM,exp}$). Upper/lower rows show Type-I/II Higgs couplings to SM fermions; columns represent BSM fermion-Higgs coupling configurations.}
\label{fig:scanSD2HDM}
\end{figure}

Figure~\ref{fig:SD2HDMblind} displays DM-proton scattering cross-sections with LZ constraints for various Yukawa configurations. Our parameter scan covers:
\begin{align}
m_S &\in [1,5000],\text{GeV}, \quad m_D \in [1,5000],\text{GeV}, \quad y \in [10^{-3},10], \nonumber\\
\tan\theta &\in [-20,20], \quad \tan\beta \in , \quad m_A \in [10,1500],\text{GeV}, \nonumber\\
m_H &\in [m_h,1500],\text{GeV}, \quad m_{H^\pm} \in [m_W,1500],\text{GeV},
\end{align}
considering scalar potential stability, perturbative unitarity, Z/h invisible decays, and relic density constraints. Results in Fig.~\ref{fig:scanSD2HDM} show rescaled cross-sections $\xi\sigma_{\chi p}^{\rm SI}$ ($\xi \equiv \Omega_{\chi}/\Omega_{\rm DM,exp}$) versus DM mass for all viable points.

The scan has been repeated for each configuration of the Yukawa couplings of the BSM fermions and for the Type-I and Type-II configurations of the Yukawa couplings of the SM fermions. Even if, similarly to the minimal singlet-doublet model, the majority of the model points are already ruled-out or in the reach of next future detectors, a sizable portion lies well inside the neutrino floor. As already commented e.g. in \cite{Arcadi:2024ukq}, the reason relies on the fact that the interactions of the DM with the pseudoscalar Higgs allow for efficient DM annihilations processes while the impact on the DM scattering processes is more limited has they enter only in loop contributions. A full computation of eq. \ref{eq:SD_2HDM_full_loop} it is then crucial to assess the detection prospects of the singlet-doublet+2HDM.

\section{2HDM+$a$}\label{sec:mod2}

The 2HDM+$a$ is a widely investigated benchmark model for both dark matter (DM) and collider searches; see, e.g., \cite{Ipek:2014gua,Bauer:2017ota,Arcadi:2020gge,Robens:2021lov,Arcadi:2022lpp,Argyropoulos:2022ezr,Argyropoulos:2024yxo}. 
The model is characterized by a Higgs sector comprising two $SU(2)_L$ doublets, $\Phi_{1,2}$, and a pseudoscalar gauge singlet $a^0$, with interactions encoded in the scalar potential:
\begin{align}
    & V_{\rm 2HDMa}=V_{\rm 2HDM}+V_{a^0},\quad \text{where} \nonumber\\
    & V_{a^0}=\frac{1}{2} m_{a^0}^2 (a^0)^2+ \frac{\lambda_a}{4} (a^0)^4+ \left(i \kappa a^0 \Phi^{\dagger}_1\Phi_2+\text{H.c.}\right)
    + \left[\lambda_{1P}(a^0)^2 \Phi_1^{\dagger}\Phi_1 + \lambda_{2P}(a^0)^2 \Phi_2^{\dagger}\Phi_2\right],\label{eq:V2HDMa}
\end{align}
After electroweak symmetry breaking, the spectrum contains two CP-even neutral scalars $h$ and $H$ (with $h$ identified as the $125$ GeV SM-like Higgs), one charged Higgs $H^{\pm}$, and two CP-odd neutral pseudoscalars $a$ and $A$, where (unless otherwise specified) we assume $m_a < m_A$. The two CP-odd states are mixtures of the singlet and doublet components, governed by a mixing angle $\theta$:
\begin{equation}
    \tan 2\theta = \frac{2 \kappa v_h}{m_A^2 - m_a^2}.
\end{equation}

The physical Higgs bosons couple to SM fermions as well as to a fermionic Dirac DM candidate $\chi$, a gauge singlet, through the Yukawa Lagrangian:
\begin{align}
-{\cal L}_{\rm Yuk} &= \sum\limits_{f=u,d,l} \frac{m_f}{v_h} \left[g_{hff} \bar{f}f h + g_{Hff} \bar{f}f H - i g_{Aff} \bar{f} \gamma_5 f A \right] \notag \\
&\quad - \frac{\sqrt{2}}{v_h} \left[ \bar{u} \left(m_u g_{Auu} P_L + m_d g_{Add} P_R \right)d H^+ + m_l g_{All} \bar{\nu} P_R \ell H^+ + \text{H.c.} \right]\notag\\
&\quad + i y_\chi \bar{\chi} \gamma_5 \chi \left(a \cos\theta + A \sin\theta\right),
\end{align}
where the Higgs-fermion couplings are expressed in terms of SM-like couplings multiplied by scaling factors $g_{\phi ff}$. For sizeable $\sin\theta$, both pseudoscalar states contribute to DM annihilation, providing a ``double portal'' to the fermionic DM, enabling consistency with the thermal freeze-out paradigm for appropriate values of $y_\chi$.

In contrast to the singlet-doublet model, tree-level DM–nucleon scattering is naturally suppressed here: the CP-odd couplings yield spin-independent (SI) interactions that vanish in the non-relativistic limit due to momentum suppression. As a result, loop effects are crucial to evaluate the direct detection (DD) phenomenology. The one-loop SI cross-section receives contributions from triangle and box diagrams analogous to those appearing in singlet-doublet + 2HDM frameworks. It reads:
\begin{align}
\label{eq:2HDMa_full_loop}
\sigma_{\chi_1^0 p}^{\rm SI} &= \frac{\mu_{\chi_1^0 p}^2}{\pi}\frac{m_p^2}{v_h^2} \left| \sum_q f_q \sum_{\phi=h,H}\frac{g_{\phi qq} m_q}{v_h m_\phi^2} C_{q, \rm triangle}
+ \sum_{q=u,d,s} f_q C_{q,\rm box} \right. \notag\\
&\qquad\left. + \sum_{q=u,d,s,c,b} \frac{3}{4}(q(2)+\bar q (2))\left[C_{1,\rm box}+m_{\chi_1^0}C_{2,\rm box}\right]
+ \frac{2}{27}f_{TG}C_{G,\rm box} \right|^2.
\end{align}

\begin{figure}
    \centering
    \subfloat{\includegraphics[width=0.27\linewidth]{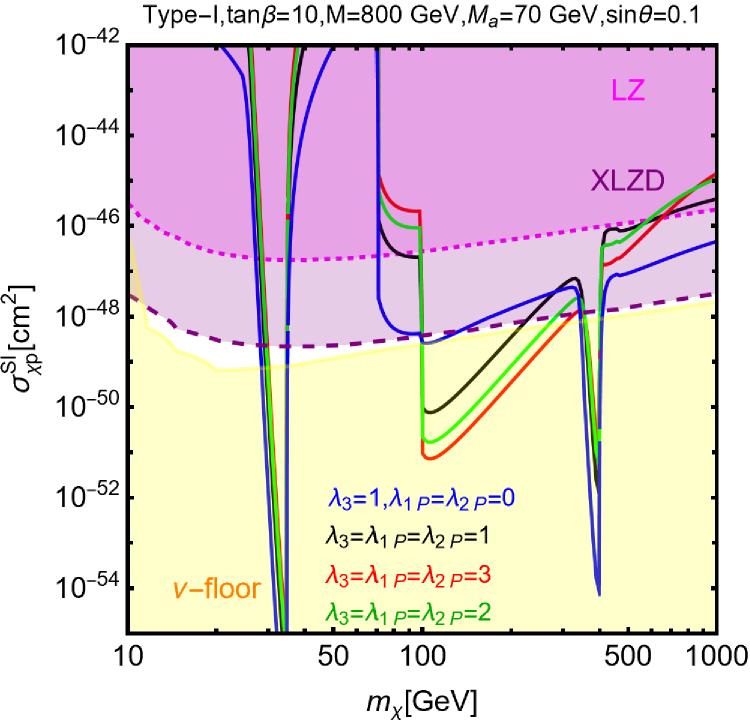}}
    \subfloat{\includegraphics[width=0.27\linewidth]{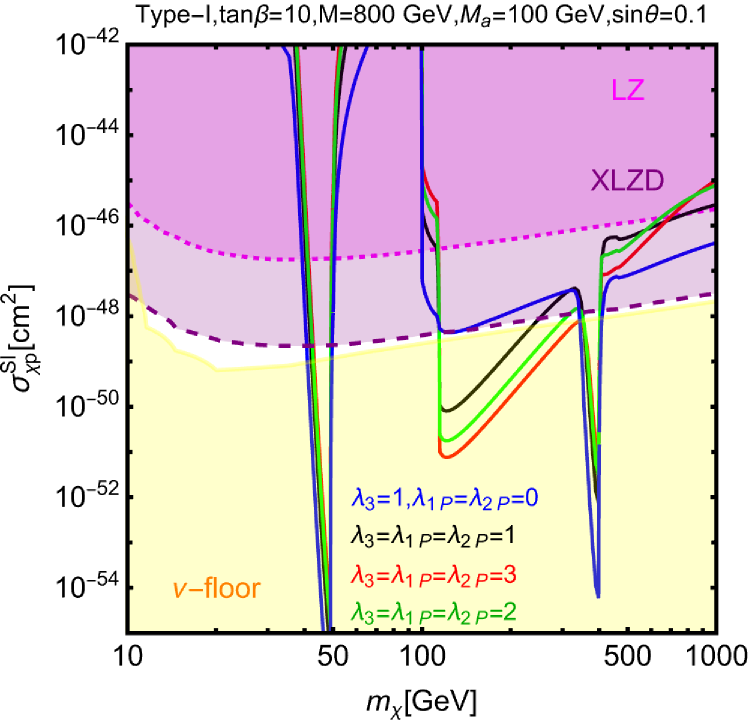}}
    \subfloat{\includegraphics[width=0.27\linewidth]{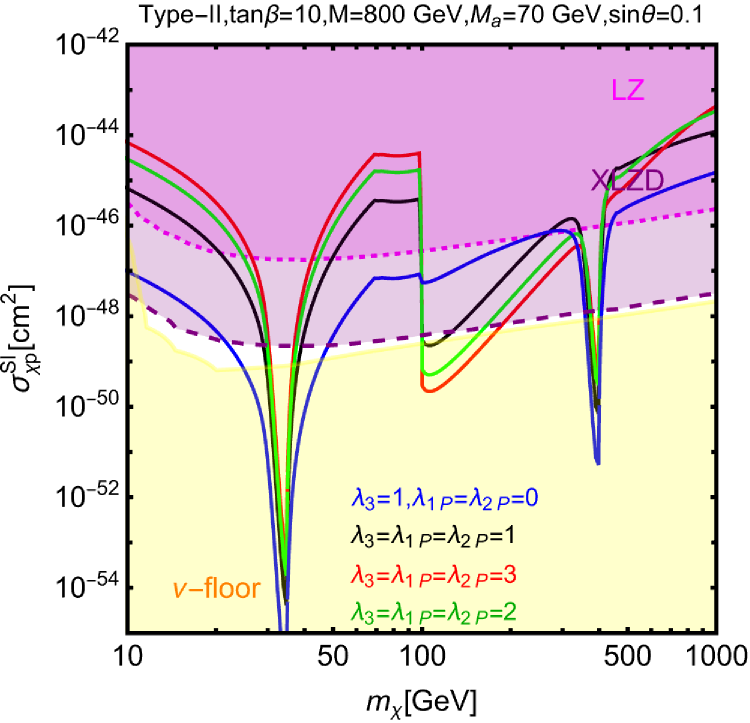}}
    \subfloat{\includegraphics[width=0.27\linewidth]{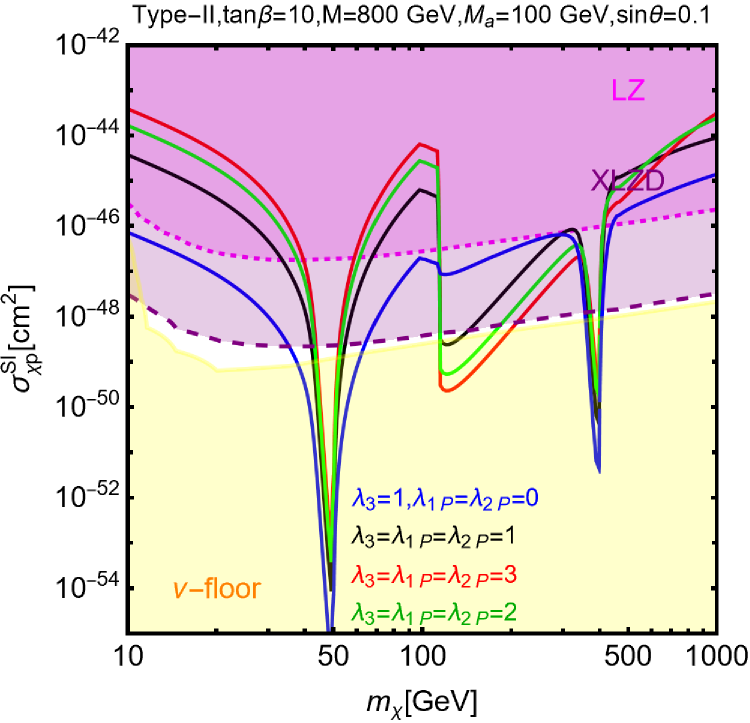}}\\
    \subfloat{\includegraphics[width=0.27\linewidth]{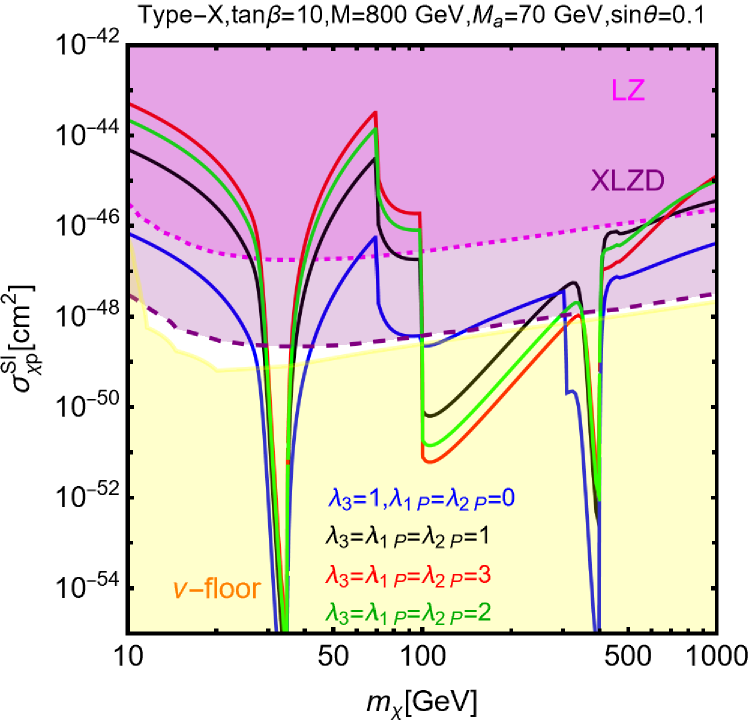}}
    \subfloat{\includegraphics[width=0.27\linewidth]{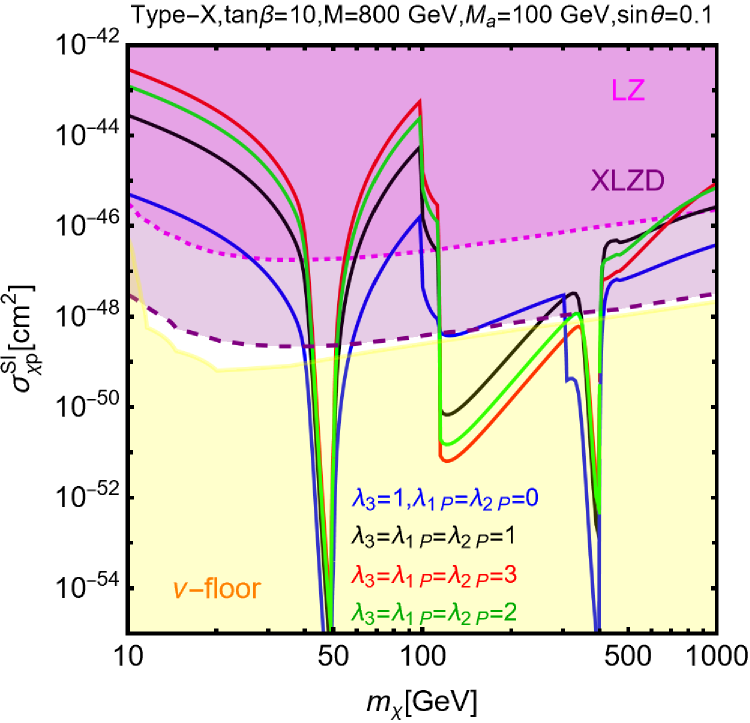}}
    \subfloat{\includegraphics[width=0.27\linewidth]{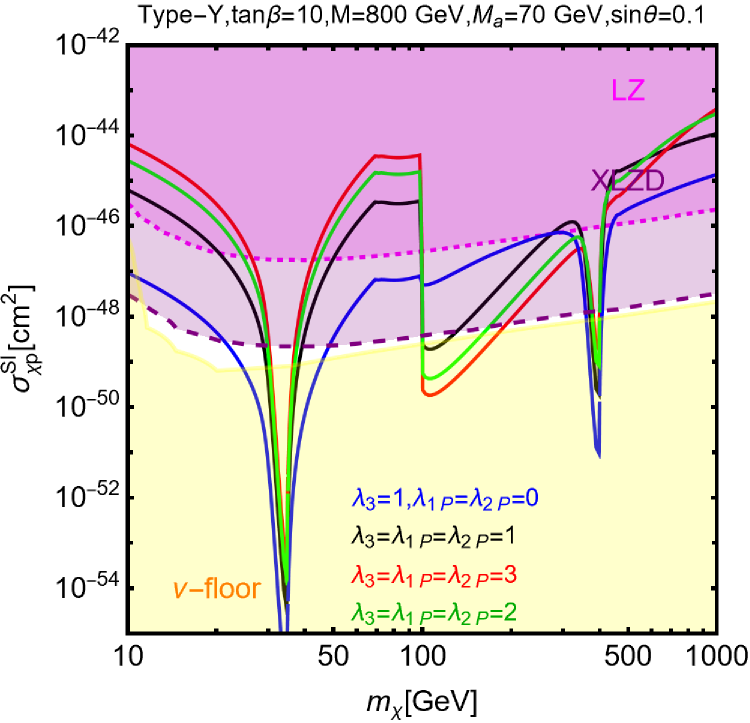}}
    \subfloat{\includegraphics[width=0.27\linewidth]{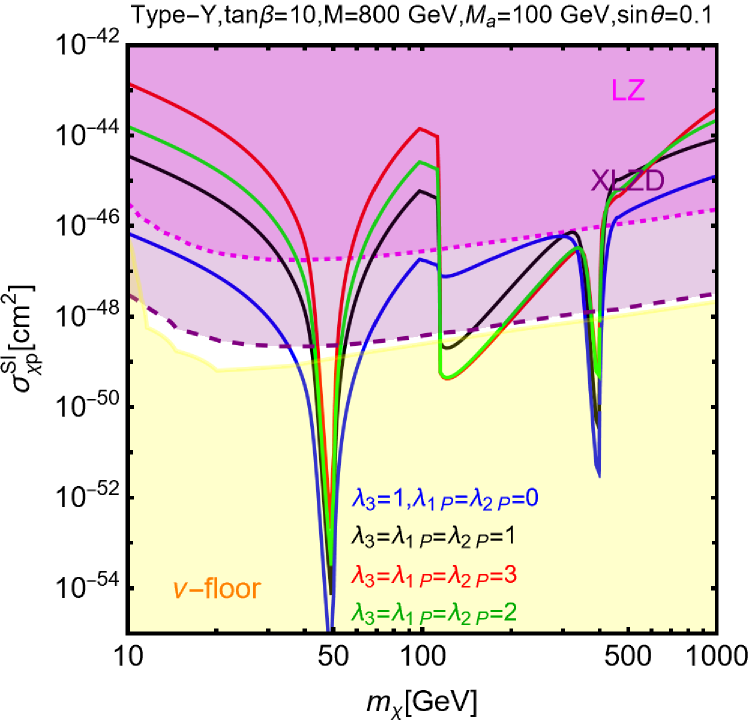}}
    \caption{\footnotesize{SI DM scattering cross-section off protons as a function of the DM mass, for representative benchmark parameter sets indicated above each panel. The solid lines correspond to relic density values consistent with thermal freeze-out, obtained by fixing $y_\chi$ accordingly. Different curves refer to varying trilinear couplings $\lambda_3$, $\lambda_{1P}$, and $\lambda_{2P}$. Each row corresponds to a different BSM Yukawa coupling type: Type-I, Type-II, Type-X, and Type-Y.}}
    \label{fig:2HDMa2D}
\end{figure}

The interplay between relic density and DD constraints is illustrated in Fig.~\ref{fig:2HDMa2D}. Each panel shows contours of the predicted DM–nucleon scattering cross-section versus DM mass for selected values of the scalar quartic couplings $\lambda_3$, $\lambda_{1P}$, and $\lambda_{2P}$, and fixed inputs for $\tan\beta$, $m_a$, $\sin\theta$, and $M \equiv m_H = m_A = m_{H^\pm}$. The BSM Yukawa coupling types vary across the panels. For each point, the coupling $y_\chi$ is numerically determined to yield the correct DM relic density.

\begin{figure}
    \centering
    \subfloat{\includegraphics[width=0.28\linewidth]{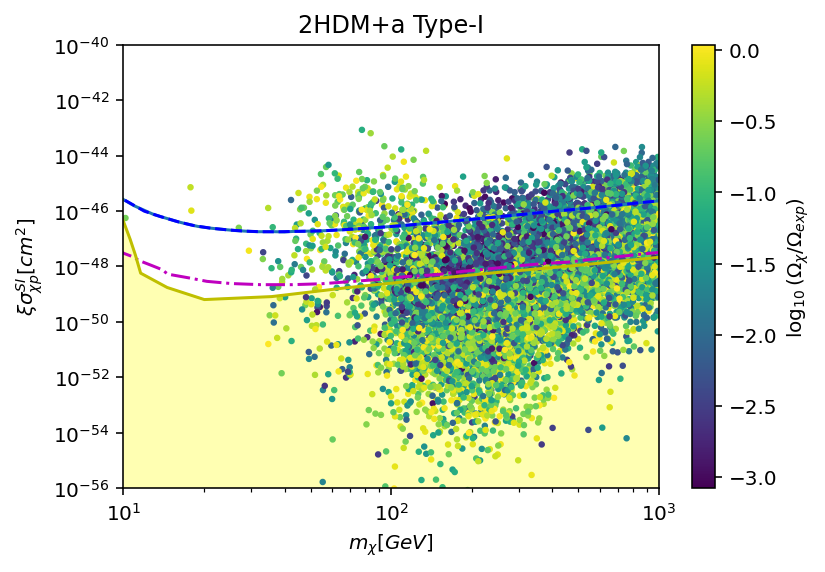}}
    \subfloat{\includegraphics[width=0.28\linewidth]{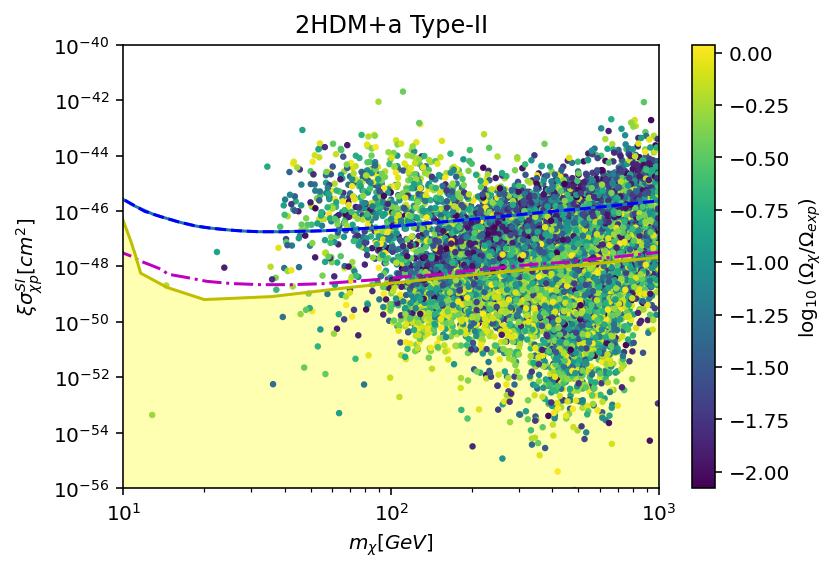}}
    \subfloat{\includegraphics[width=0.28\linewidth]{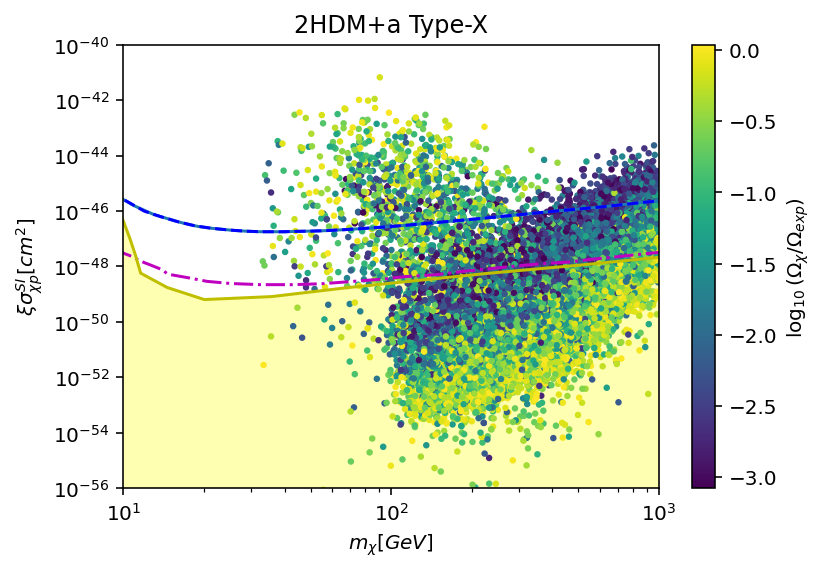}}
    \subfloat{\includegraphics[width=0.28\linewidth]{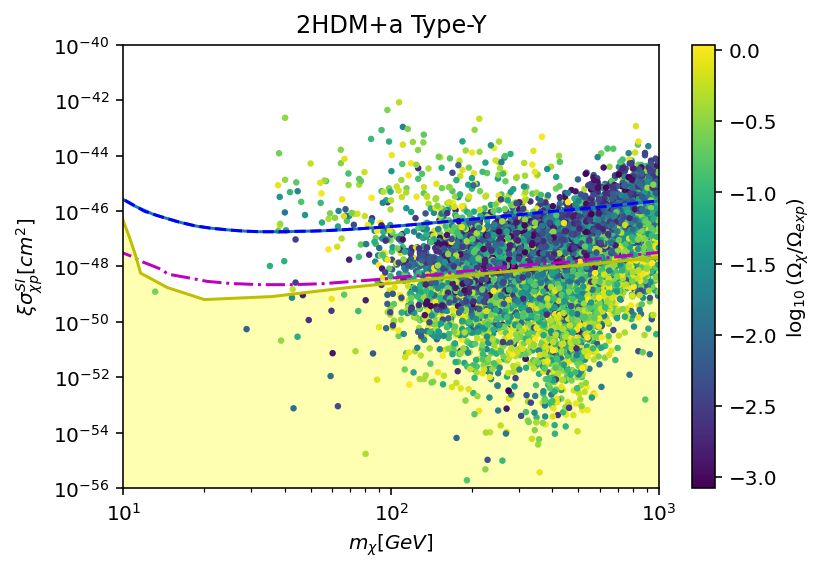}}\\
    \subfloat{\includegraphics[width=0.28\linewidth]{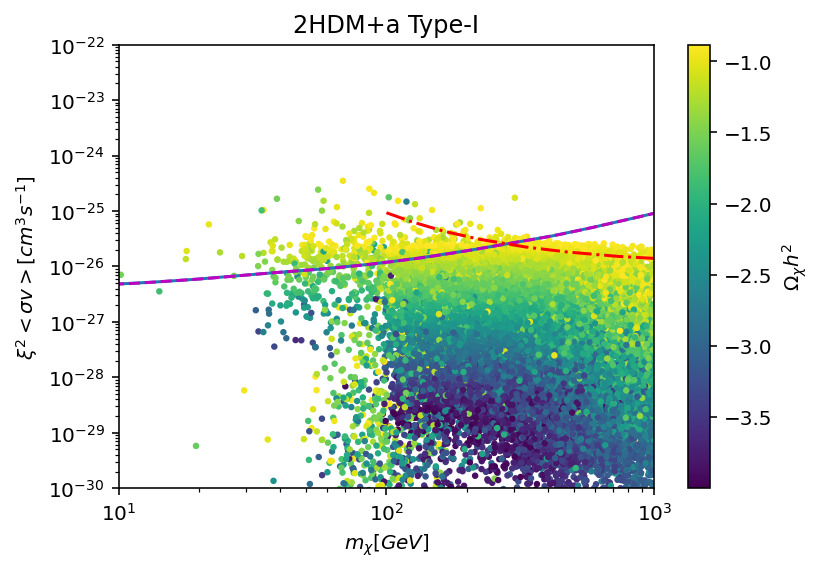}}
    \subfloat{\includegraphics[width=0.28\linewidth]{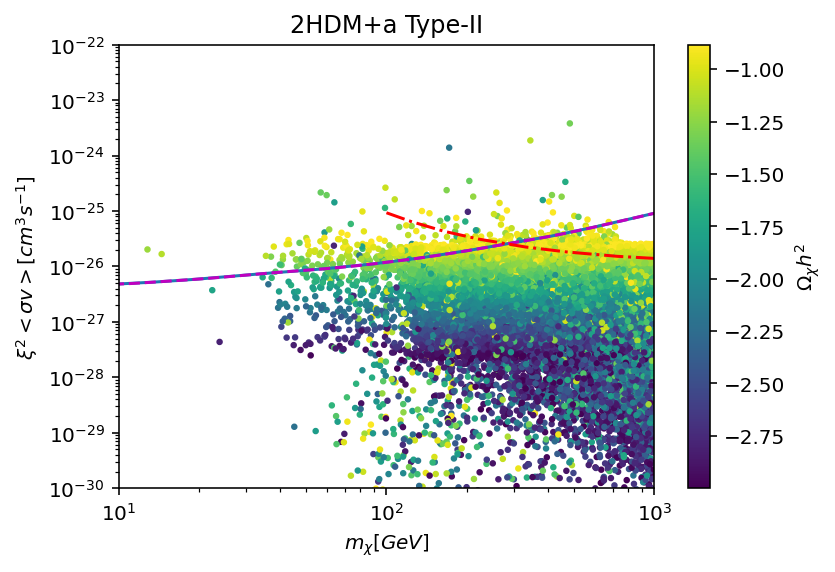}}
    \subfloat{\includegraphics[width=0.28\linewidth]{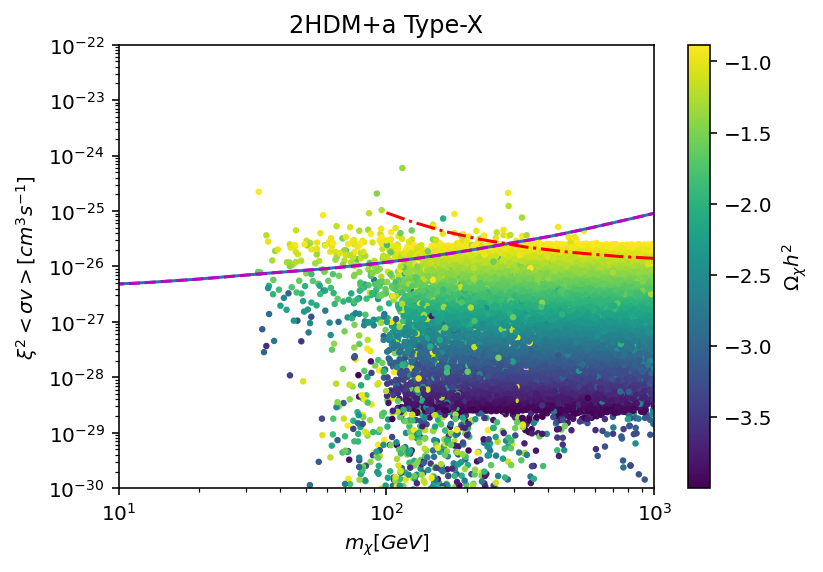}}
    \subfloat{\includegraphics[width=0.28\linewidth]{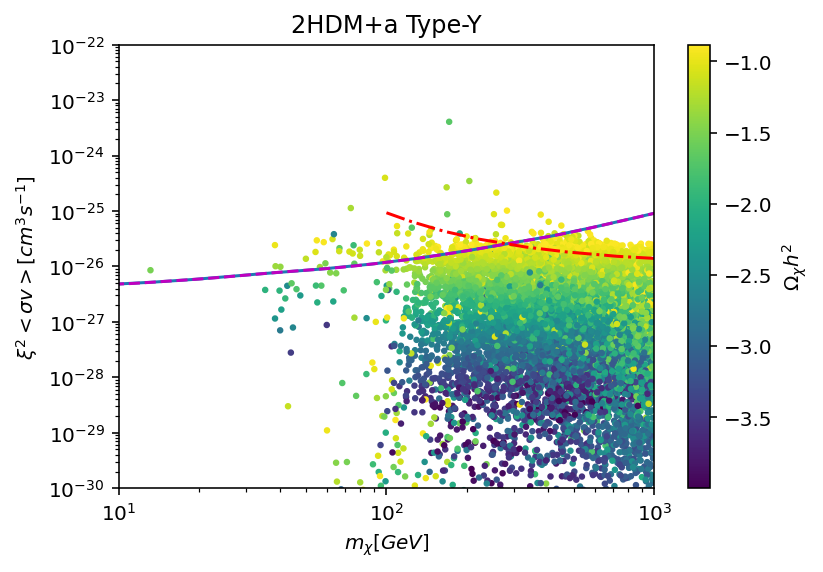}}
    \caption{\footnotesize{Model points in the $(m_\chi,\xi \sigma_{\chi p}^{\rm SI})$ plane from the scan described in Eq.~\eqref{eq:2HDMascan}. The color scale denotes the relic density, which is subdominant in all points. Scattering cross-sections are rescaled by $\xi = \Omega_\chi / \Omega_{\rm DM,exp}$. Panels correspond to Type-I, II, X, and Y Yukawa structures, respectively.}}
    \label{fig:2HDMascan}
\end{figure}

A broader parameter space exploration has been carried out via the following scan ranges:
\begin{align}
\label{eq:2HDMascan}
    & m_\chi \in [1,1000]\,\mathrm{GeV},\quad m_a \in [10,600]\,\mathrm{GeV},\notag\\
    & m_H,\, m_{H^{\pm}},\, m_A \in [m_h,1500]\,\mathrm{GeV},\quad y_\chi \in [10^{-3},10],\notag\\
    & \sin\theta \in \left[-\frac{\pi}{4},\frac{\pi}{4}\right],\quad \tan\beta \in [1,60],\notag\\
    & \lambda_3,\, \lambda_{1P},\, \lambda_{2P} \in [-4\pi,4\pi].
\end{align}
The scan has been repeated for all four Yukawa coupling structures: Type-I, Type-II, Type-X, and Type-Y. Results are summarized in Fig.~\ref{fig:2HDMascan}.

Only points satisfying constraints from vacuum stability, perturbative unitarity, flavor physics, and Higgs invisible decays (see \cite{Arcadi:2022lpp}) are retained. The top and bottom panels of Fig.~\ref{fig:2HDMascan} show, respectively, the predicted SI scattering cross-sections and present-day annihilation cross-sections, both rescaled by $\xi$ and $\xi^2$ to reflect scenarios where the fermionic DM constitutes only a subcomponent of the total relic abundance. The DM scattering cross-section has been compared, as customary with the current (projected) limits by LZ (XLZD) while, for what indirect limits are concerned, we have followed the same procedure as \cite{Arcadi:2024ukq} and considered the exclusion bound from DSph as given by FERMI \cite{McDaniel:2023bju} as well as the projected sensitivity by CTA \cite{CTA:2020qlo}\footnote{\footnotesize{For definiteness we have considered the $\bar b b$ final state.}}.

So far, we have considered the regime where both $y_\chi$ and $\sin\theta$ are sizeable. However, viable DM scenarios also exist in alternative regions of parameter space. In particular, \cite{Haisch:2023rqs} explored the case of highly suppressed $\sin\theta$ with large $y_\chi$, where DM annihilation proceeds predominantly into Higgs bosons. Loop-induced SI scattering remains operative due to nonzero $\lambda_{haa}$ and $\lambda_{Haa}$ even as $\sin\theta \rightarrow 0$:
\begin{align}
\label{eq:sth0}
    \lambda_{haa} &= -\frac{2v_h}{m_h}\left(\lambda_{1P}\cos^2 \beta + \lambda_{2P}\sin^2 \beta\right), \notag\\
    \lambda_{Haa} &= \frac{v_h}{m_H}\sin 2\beta \left(\lambda_{1P} - \lambda_{2P}\right).
\end{align}

\begin{figure}
    \centering
    \subfloat{\includegraphics[width=0.5\linewidth]{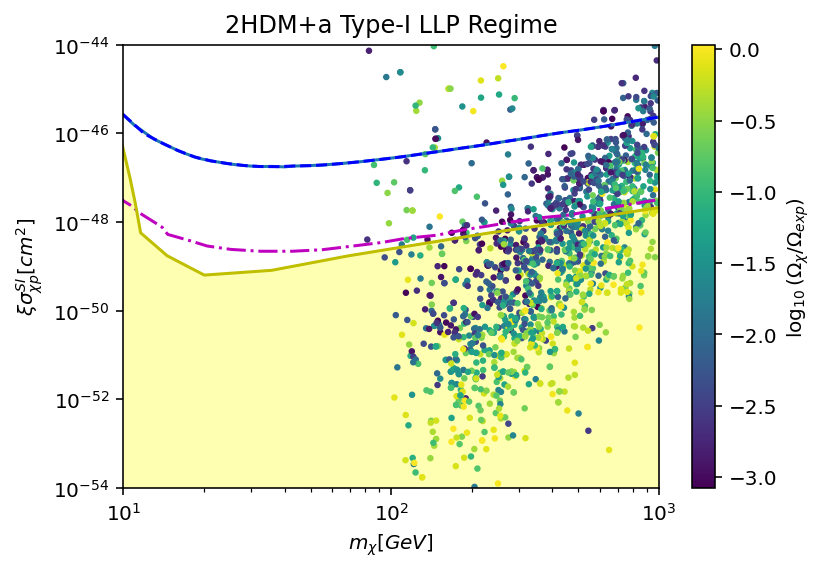}}
    \subfloat{\includegraphics[width=0.5\linewidth]{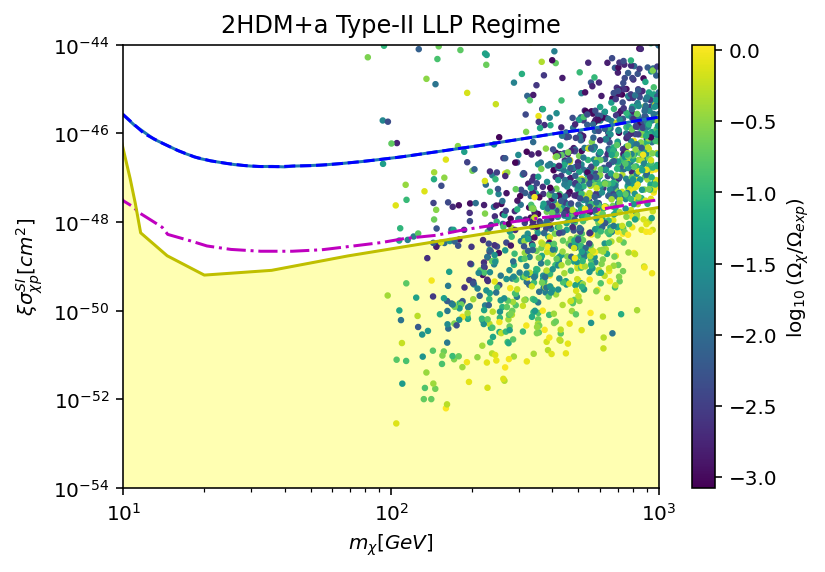}}
    \caption{\footnotesize{Results from a dedicated scan in the regime $\sin\theta \in [10^{-10},10^{-6}]$, corresponding to a collider-stable (long-lived) light pseudoscalar $a$. Panels correspond to Type-I and Type-II Yukawa scenarios.}}
    \label{fig:p2HDMaLLP}
\end{figure}

To explore this regime, we repeated the scan with $\sin\theta \in [10^{-10},10^{-6}]$ (see Fig.~\ref{fig:p2HDMaLLP}). For simplicity, only the Type-I and Type-II structures are shown. In this limit, viable DM masses are generally above 100 GeV, and most points satisfying relic density constraints lie within the so-called neutrino floor, limiting the prospects for direct detection.

Finally, we note the intriguing possibility of setting $\sin\theta=0$ exactly. If $m_a < m_\chi$, the $a$ field can then be a cosmologically stable pseudoscalar that is entirely a gauge singlet but remains coupled to the SM via $\lambda_{1P}$ and $\lambda_{2P}$. In such a scenario, a two-component DM framework naturally arises, composed of $\chi$ and $a$. Two-component fermion–pseudoscalar simplified models have previously been considered in the literature; see, e.g., \cite{DiazSaez:2021pmg,Belyaev:2022qnf}.




\section{Dark \texorpdfstring{$SU(3)$}{SU(3)} Model}\label{sec:mod3}

The Dark $SU(3)$ model extends the Standard Model (SM) by introducing a ``dark'' $SU(3)$ gauge symmetry, which is completely spontaneously broken via two $SU(3)$ triplet scalars with misaligned vacuum expectation values (vevs). The dark sector communicates with the SM through mass mixing between the dark and SM Higgs bosons. Upon symmetry breaking, a discrete remnant symmetry remains, stabilizing the lightest massive dark gauge boson. In addition, if CP is conserved in the scalar sector, a pseudoscalar mass eigenstate $\Psi$ from the Higgs sector can also be cosmologically stable.

The model is detailed in Refs.~\cite{Gross:2015cwa,Arcadi:2016kmk}; we therefore omit a complete description here. Assuming a strong hierarchy in the vevs of the $SU(3)$ triplets and appropriate choices of the scalar potential parameters, the subset of particles relevant for dark matter (DM) phenomenology is governed by the Lagrangian:
\begin{align}
\label{eq:SU3_lagrangian}
     \mathcal{L} &= \frac{\tilde{g} m_{V}}{2} \left(-\sin\theta\, H_1 + \cos\theta\, H_2\right) \left( \sum_{a=1}^{2} V_{\mu}^a V^{\mu\,a} + \left(\cos\alpha - \frac{\sin\alpha}{\sqrt{3}}\right)^2 V_\mu^3 V^{\mu\,3} \right)\nonumber\\
    & + \tilde{g}\cos\alpha \sum_{a,b,c} \epsilon_{abc} \partial_\mu V_\nu^a V^{b\,\mu} V^{c\,\nu}
    - \frac{\tilde{g}^2}{2} \cos^2\alpha \sum_{a=1}^{2} \left( V_\mu^a V^{a\,\mu} V_\nu^3 V^{3\,\nu} - \left(V_\mu^a V^{a\,\mu}\right)^2 \right) \nonumber\\
    & + \left[ \frac{\tilde{g}}{2 M_V} \left(-\sin\theta\, m_{H_1}^2 H_1 + \cos\theta\, m_{H_2}^2 H_2\right) - \frac{1}{4} \left( \lambda_{\psi\psi 11} H_1^2 + 2 \lambda_{\psi\psi 12} H_1 H_2 + \lambda_{\psi\psi 22} H_2^2 \right) \right] \psi^2 \nonumber\\
    & - \frac{\kappa_{111}}{2} v_h H_1^3 
    - \frac{\kappa_{112}}{2} H_1^2 H_2 v_h \sin\theta 
    - \frac{\kappa_{221}}{2} H_1 H_2^2 v_h \cos\theta 
    - \frac{\kappa_{222}}{2} H_2^3 v_h,
\end{align}
where $V^{1,2}$ denote a pair of mass-degenerate vector fields of mass $m_V$, representing one DM component. The second component is the lightest among $V^3$, a third dark $SU(3)$ vector boson with mass $m_V (\cos \alpha - \sin \alpha/\sqrt{3})^2 \simeq m_V$ (since $\sin \alpha \ll 1$), and the pseudoscalar state $\Psi$ from the scalar sector.\footnote{This two-component setup is realized only if CP symmetry is conserved in the scalar sector.}

The states $H_1$ and $H_2$ are two Higgs mass eigenstates, with $H_1$ identified as the SM-like Higgs at 125 GeV. They are linear combinations of the SM and dark sector Higgs fields, mixed by an angle $\theta$.\footnote{The remaining Higgs states are assumed to be heavy and weakly coupled to the SM.}

In general, the DM relic abundance in this two-component DM (2CDM) framework must be obtained by solving a coupled system of Boltzmann equations. In the vector/vector case, and in the $\sin\alpha \ll 1$ regime, the vector components have nearly identical masses and couplings. Moreover, as shown in Ref.~\cite{Gross:2015cwa}, semi-annihilation processes of the form $VV \to V H_{1,2}$ are subdominant. Thus, the relic density calculation closely resembles the standard single-component freeze-out scenario.

In contrast, in scalar/vector scenarios, the two components may differ significantly in mass and interactions. The coupled Boltzmann equations for this case are:
\begin{align}
    \frac{dY_V}{dx} &= -\frac{s}{Hx} \left[ \langle \sigma v \rangle_{VV \to XX} \left(Y_V^2 - Y_{V,eq}^2\right)
    + \langle \sigma v \rangle_{VV \to \Psi\Psi} \left(Y_V^2 - \frac{Y_{V,eq}^2}{Y_{\Psi,eq}^2} Y_\Psi^2\right) \right. \nonumber\\
    & \left. + \langle \sigma v \rangle_{VV \to V^3 H_{1,2}} \left(Y_V^2 - \frac{Y_{\Psi,eq}}{Y_\Psi} Y_V^2 \right) \right], \nonumber\\
    \frac{dY_\Psi}{dx} &= -\frac{s}{Hx} \left[ \langle \sigma v \rangle_{\Psi\Psi \to XX} \left(Y_\Psi^2 - Y_{\Psi,eq}^2\right) 
    - \langle \sigma v \rangle_{V V^3 \to V H_{1,2}} Y_V Y_{V^3,eq} \left(\frac{Y_\Psi}{Y_{\Psi,eq}} - 1 \right) \right] \nonumber\\
    & + \frac{s}{Hx} \left[ \langle \sigma v \rangle_{VV \to \Psi\Psi} \left(Y_V^2 - \frac{Y_{V,eq}^2}{Y_{\Psi,eq}^2} Y_\Psi^2 \right) 
    + \langle \sigma v \rangle_{VV \to V^3 H_{1,2}} \left(Y_V^2 - \frac{Y_{\Psi,eq}}{Y_\Psi} Y_V^2 \right) \right],
\end{align}
where $x \equiv m_V/T$, and $X$ denotes either a SM field or a Higgs boson $H_{1,2}$. Semi-annihilations and DM interconversions are crucial to determine the correct relic abundance~\cite{Arcadi:2016kmk}.

\begin{figure}
    \centering
    \includegraphics[width=0.4\linewidth]{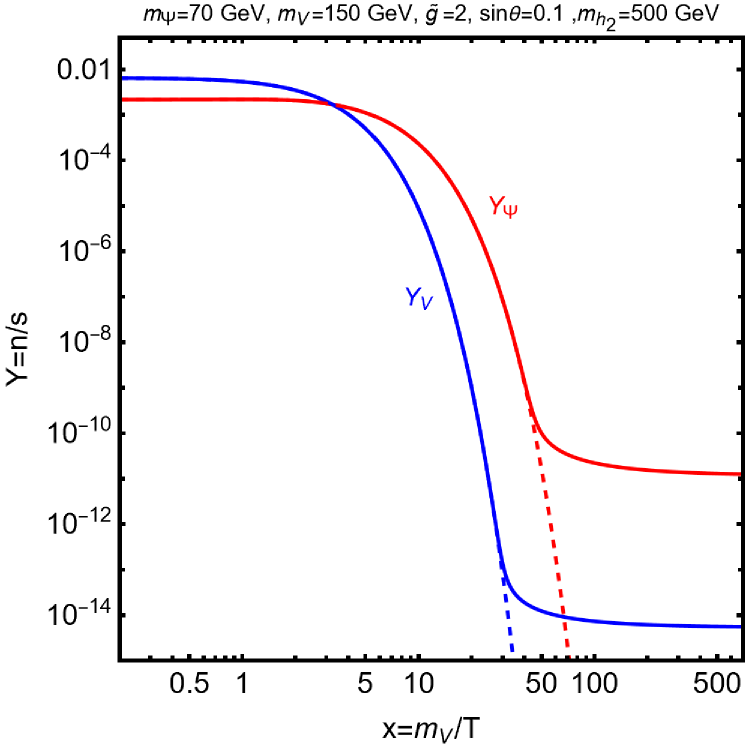}
    \includegraphics[width=0.4\linewidth]{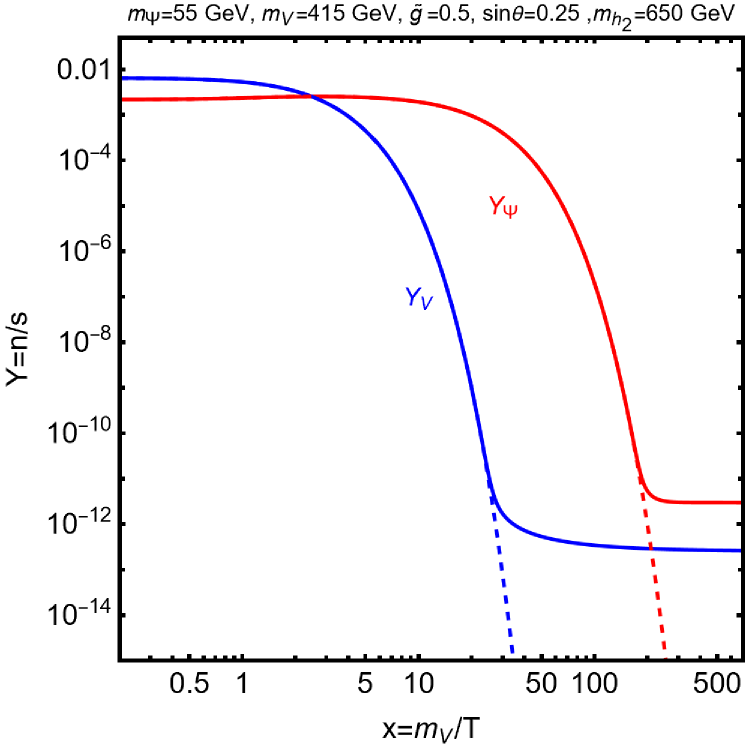}\\
    \includegraphics[width=0.4\linewidth]{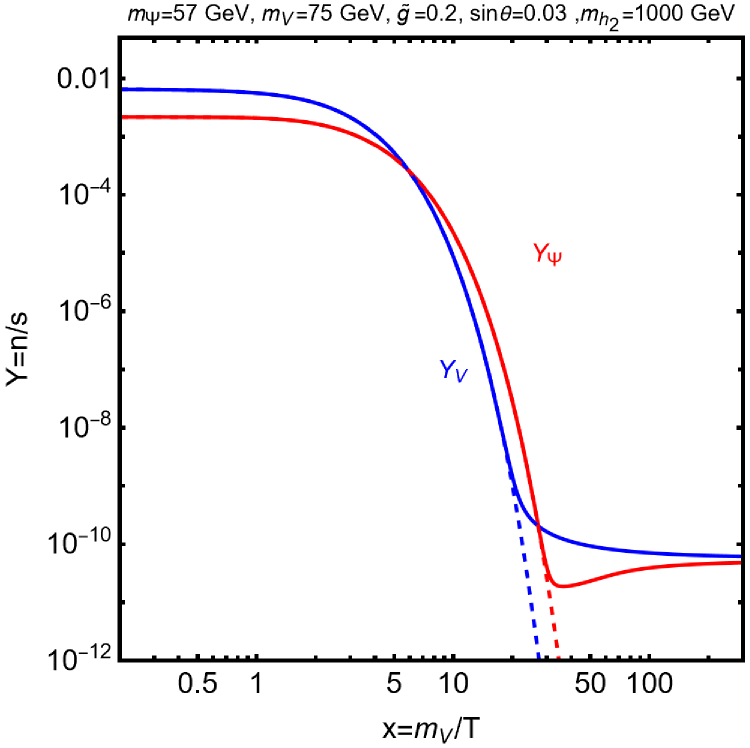}
    \caption{\footnotesize Evolution of the comoving number densities of the scalar ($\Psi$) and vector ($V$) DM components in the dark $SU(3)$ model for various benchmark parameter choices, reported atop each panel.}
    \label{fig:pbolSU3}
\end{figure}

As shown in Fig.~\ref{fig:pbolSU3}, the Boltzmann system has been solved numerically for benchmark parameter values $(\tilde{g}, M_{H_2}, \sin\theta)$ in the scalar/vector scenario.

An interesting feature of this model is that the scalar DM component $\Psi$ completely evades direct detection (DD). Its coupling to the Higgs states is:
\begin{equation}
    g_{\psi\psi H_i} = \frac{\tilde{g}}{2 m_V^2} \delta_i m_{H_i}^2,
\end{equation}
with $\delta_i = \sin\theta$ ($\cos\theta$) for $H_1$ ($H_2$). The spin-independent DM-nucleon cross-section is:
\begin{equation}
    \sigma_{\psi p}^{\rm SI} \propto \left| \frac{g_{\psi\psi H_1} \cos\theta}{m_{H_1}^2} + \frac{g_{\psi\psi H_2} \sin\theta}{m_{H_2}^2} \right|^2 = 0,
\end{equation}
due to an exact cancellation enforced by the scalar potential. This is not a ``blind spot'' dependent on parameter tuning; the cancellation holds generically across parameter space. As a result, only the vector DM component is accessible to DD experiments.

\begin{figure}
    \centering
    \subfloat{\includegraphics[width=0.6\linewidth]{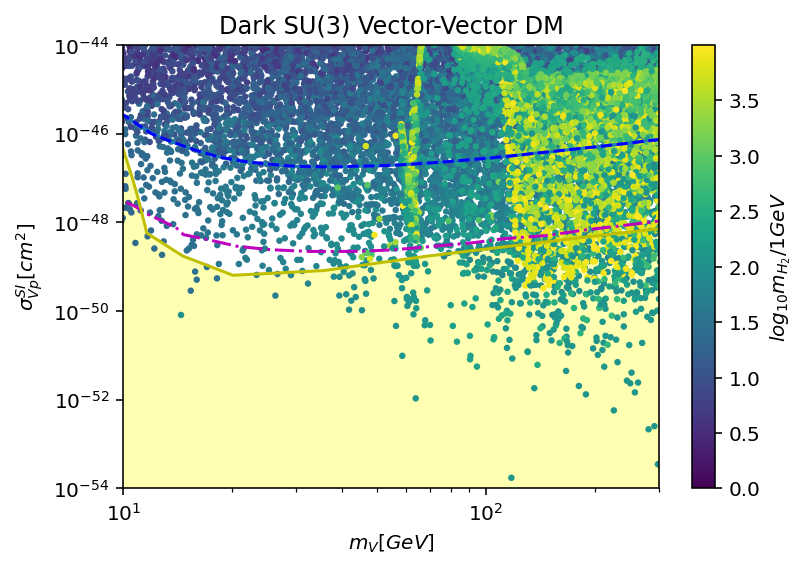}}\\
    \subfloat{\includegraphics[width=0.6\linewidth]{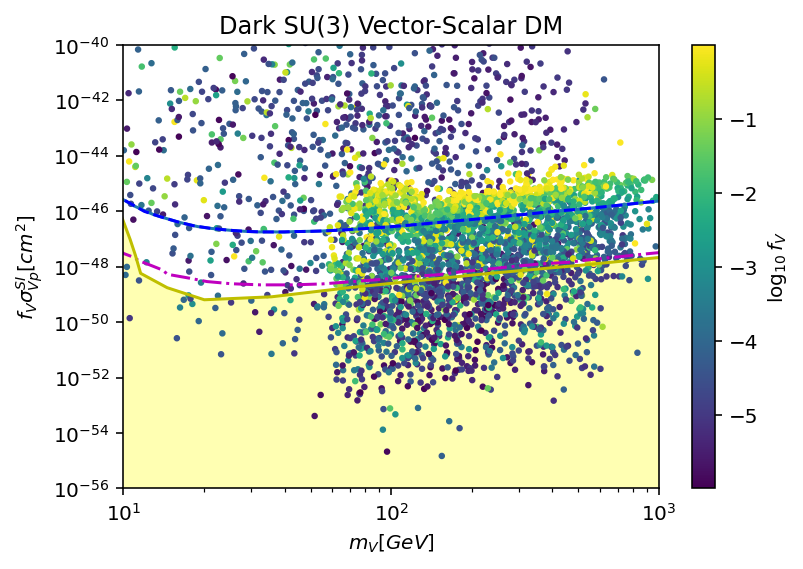}}
    \caption{\footnotesize Parameter scan results for the dark $SU(3)$ model. In both panels, the total DM abundance agrees with the observed relic density within $3\sigma$. Top: vector/vector scenario with color encoding $M_{H_2}$. Bottom: scalar/vector case, showing the vector DM scattering cross-section on protons, rescaled by the relic fraction $f_V$.}
    \label{fig:darkSU3scan}
\end{figure}

Figure~\ref{fig:darkSU3scan} presents the outcome of a parameter scan over $(\tilde{g}, \sin\theta, M_{H_2})$, taken in the ranges:
\begin{equation}
    \tilde{g} \in [10^{-2}, 10], \quad \sin\theta \in [10^{-3}, 0.707], \quad M_{H_2} \in [1, 1000]\,\text{GeV}.
\end{equation}
For the vector/vector setup, $m_\Psi = 300$ GeV was fixed, and $m_V \in [1, 300]$ GeV. For scalar/vector scenarios, $m_\Psi \in [1, 300]$ GeV and $m_V \in [1\,\text{GeV}, 1\,\text{TeV}]$, with the requirement $m_\Psi < m_V$. Only points consistent with the observed relic density are retained.

Both panels show results in the $(m_V, \sigma_{Vp}^{\rm SI})$ plane. In the vector/vector case, the two components are nearly degenerate and behave as a single DM species for DD. In the scalar/vector setup, the color scale indicates the fractional contribution $f_V$ of the vector component to the total relic density.

The results show that the vector/vector model, like other WIMP setups, will be thoroughly probed by upcoming DD experiments. In contrast, detection prospects in the scalar/vector scenario hinge solely on the vector component, whose relic fraction may be subdominant. Nonetheless, next-generation DD experiments will be sensitive to scenarios with $f_V \gtrsim 10^{-3}$.

\newpage\clearpage

\section{Discussion and Conclusions}\label{sec:conclusions}

In this work, we provided a detailed assessment of three distinct classes of dark matter models that exemplify scenarios in which the spin-independent elastic scattering cross-section of WIMPs with nucleons is strongly suppressed. Each model framework—namely, the minimal singlet-doublet model, its extension with an additional two-Higgs-doublet plus pseudoscalar scalar sector, and a dark $SU(3)$ gauge model—serves to highlight complementary mechanisms capable of concealing WIMPs from direct detection efforts while preserving consistency with thermal relic abundance requirements.

In the case of the singlet-doublet model, we demonstrated the presence of ``blind spots’’ in parameter space where the effective coupling to the Higgs boson vanishes at tree level. Although such cancellations are not protected by symmetries and hence subject to quantum corrections, our loop-level analysis shows that viable parameter regions remain, with suppressed cross-sections persisting even after the inclusion of radiative effects. The addition of a second Higgs doublet and a pseudoscalar mediator enriches the phenomenology and opens up new possibilities for suppressing the direct detection signal through destructive interference between multiple CP-even Higgs exchange diagrams. Our scan over various Yukawa coupling assignments reveals that this extended setup can accommodate sizable portions of viable parameter space where the WIMP relic density is correct and direct detection limits are evaded.

The 2HDM$+a$ model, extensively studied in both dark matter and collider contexts, naturally suppresses spin-independent scattering due to the pseudoscalar nature of the mediator. As such, tree-level contributions to direct detection vanish in the non-relativistic limit, and the relevant constraints arise exclusively from loop-induced effects. We performed a dedicated computation of these contributions, identifying parameter regions where the correct relic density is obtained via annihilation through the mixed pseudoscalar portal, yet direct detection remains elusive. Furthermore, we examined the limit of small mixing between the pseudoscalar mediator and the Higgs doublets, revealing the emergence of a two-component dark matter scenario when the light pseudoscalar becomes stable.

The dark $SU(3)$ model provides an inherently multi-component framework, with stability emerging from residual symmetries in the broken gauge group. Remarkably, one of the dark matter components—a CP-odd scalar—exhibits an exact cancellation in its coupling to nucleons, rendering it completely invisible to direct detection. This mechanism does not rely on fine-tuning or accidental cancellations and is therefore robust across parameter space. Consequently, only the vector component of dark matter contributes to the direct detection signal, and our analysis shows that, depending on its relic abundance fraction, detection prospects may still remain within reach of next-generation experiments.

Across all models considered, we emphasize that significant and well-motivated regions of parameter space persist between current bounds and the so-called neutrino floor. These scenarios are not only consistent with thermal freeze-out but also provide concrete benchmarks for future experimental searches. The continued development of ultra-sensitive detectors, along with complementary collider and astrophysical probes, will be essential to further test these elusive but theoretically compelling dark matter candidates.

\section*{Acknowledgments}

The authors warmly thank Manfred Lindner for the valuable suggestions on the draft and the fruitful discussions.
G.A. acknowledges support from the DAAD German Academic Exchange Service. G.A. thanks the MPIK for the warm hospitality during part of the completion of this work. G.A. thanks David Cabo Almeida for the valuable feedback. This work is partly supported by the U.S.\ Department of Energy grant number de-sc0010107 (SP). 

\appendix

\section{Loop functions for the minimal singlet-doublet model}
\label{sec:loop_CP_even}

In this appendix we provide the expressions of the loop induced coefficients to the SI scattering cross-section in the minimal singlet-doublet model. As already pointed out the computation is based on the results of \cite{Ertas:2019dew}.
The contribution from the Feynman's diagrams with triangle topology, corresponding to the second term in eq. \ref{eq:SD_full_loop}, is given by:
\begin{align}
    & C_q^{\rm triangle}=\frac{m_q}{v_h}\frac{m_\chi}{(4\pi)^2}\sum_{j=1,3}y_{h \chi_1^0\chi_j^0}y_{h\chi_j^0 \chi_1^0}^{*}
   \left(2 C_0 (m_\chi^2,m_{\chi_j}^2,m_h^2)+ C_2 (m_\chi^2,m_{\chi_j}^2,m_h^2)\right)
\end{align}
with $m_q$ being the quark mass while $C_0$ and $C_2$ are Passarino-Veltman functions \cite{tHooft:1978jhc,Passarino:1978jh} defined as:
\begin{align}
    \int \frac{d^4k}{(2\pi)^4}\frac{1}{\left[(p+k)^2-M^2\right]\left(k^2-m^2\right)^2}=\frac{i}{(4\pi)^2}C_0 (p^2,m^2,M^2)\nonumber\\
    \int \frac{d^4k}{(2\pi)^4}\frac{k^\mu}{\left[(p+k)^2-M^2\right]\left(k^2-m^2\right)^2}=\frac{i}{(4\pi)^2}p^\mu C_2(p^2,m^2,M^2)
\end{align}

The latter functions, together with the ones presented below and in the next appendix have been computed via the Package-X package \cite{Patel:2015tea,Patel:2016fam}.
$y_{\chi_1^0 \chi_j^0 h}$ are finally the couplings between the Higgs bosons and the BSM fermions:
\begin{equation}
y_{h N_i N_j}=\frac{1}{\sqrt{2}}\left(y_1 U_{i2}^{*}U_{j1}^{*}+y_2 U_{j3}^{*}U_{i1}^{*}\right),
\end{equation}
The analytical expression of the coupling of the SM Higgs with a DM pair has been provided in the main text.
The coefficients associated, instead, to box-shaped diagrams are given by:

\begin{align}
    & C_{1,\rm box}^{\rm CP-even}=\frac{m_q^2}{16\pi^2 v_h^2}\sum_{j=1,3} y_{\chi_1 \chi_j h}y_{h \chi_j \chi_1}^{*}m_{\chi_1^0}\left[m_{\chi_1^0}Z_{111}(m_{\chi_1^0}^2,m_{\chi_j^0}^2,m_h^2)-6 Z_{001}(m_{\chi_1^0}^2,m_{\chi_j^0}^2,m_h^2)\right.\nonumber\\
    & \left. +4 Y_2(m_{\chi_1^0}^2,m_{\chi_j^0}^2,0,m_h^2)-2 m_{\chi_1^0}^2 Z_{11}(m_{\chi_1^0}^2,m_{\chi_j^0}^2,m_h^2)-8 Z_{00}(m_{\chi_1^0}^2,m_{\chi_j}^2,m_h^2)+8 X_2(m_{\chi_1^0}^2,m_{\chi_j^0}^2,0,m_h^2) \right]\nonumber\\
    & C_{5,\rm box}^{\rm CP-even}=-\frac{m_q^2}{2 \pi^2 v_h^2}\sum_{j=1,3}y_{h \chi_1^0 \chi_j^0} y_{h \chi_j \chi_1}^{*}m_{\chi_1^0} Z_{001}(m_{\chi_1^0}^2,m_{\chi_j^0}^2,m_h^2)\nonumber\\
    & C_{6,\rm box}^{\rm CP-even}=\frac{m_q^2}{16 \pi^2 v_h^2}\sum_{j=1,3}y_{h\chi_1 \chi_j}y_{h \chi_j \chi_1}^{*}m_{\chi_1^0}\left[-4 Z_{111}(m_{\chi_1^0}^2,m_{\chi_j}^2,m_h^2)-8 Z_{11}(m_{\chi_1^0}^2,m_{\chi_j^0}^2,m_h^2)\right]\nonumber\\
    & C_{G}^{\rm CP-even}=\frac{1}{16\pi^2}\sum_{q=c,b,t}y_{h\chi_1 \chi_j}y_{h \chi_j \chi_1}^{*}\left[m_{\chi_1^0}F_Y (m_{\chi_1^0}^2,m_{\chi_j}^2,m_h^2,m_q^2)+2 m_{\chi_1^0}F_X(m_{\chi_1^0}^2,m_{\chi_j^0}^2,m_h^2,m_q^2)\right]
\end{align}

The loop functions $Z_{111},Z_{001},Z_{11},Z_{00}$ are defined through the following integrals:
\begin{align}
    & \int \frac{d^4k}{(2\pi)^4}\frac{k^\mu k^\nu}{\left[(p+k)^2-M^2\right]k^4 (k^2-m^2)^2}=\frac{i}{(4\pi)^2}\left(p^\mu p^\nu Z_{11}(p^2,M^2,m^2)+g^{\mu \nu} Z_{00}(p^2,M^2,m^2)\right)\nonumber\\
    & \int \frac{d^4k}{(2\pi)^4}\frac{k^\mu k^\nu k^\alpha}{\left[(p+k)^2-M^2\right]k^4 (k^2-m^2)^2}\nonumber\\
    & =\frac{i}{(4\pi)^2}\left(p^\mu p^\nu p^\alpha Z_{111}(p^2,M^2,m^2)+\left(g^{\mu \nu}p^\alpha+g^{\alpha \mu}p^\nu+g^{\nu \alpha}p^\mu\right)Z_{001}(p^2,M^2,m^2)\right)
\end{align}

The functions $F_{X,Y}$ are, instead, given by:
\begin{align}
    & F_{\Lambda=X,Y}(p^2,M^2,m^2,m_q^2)=\int_0^1 dx \left[-\frac{3}{2}\frac{\partial}{\partial m^2}\Lambda_1 \left(p^2,M^2,m^2,\frac{m_q^2}{x(1-x)}\right)\right.\nonumber\\
    & \left. \frac{m_q^2}{2}\frac{3 (1-x)x+2 (-1-x+x^2)}{x^2 (1-x)^2}\frac{\partial}{\partial m^2}\Lambda_2 \left(p^2,m_{\chi_j^0}^2,m_h^2,\frac{m_q^2}{x(1-x)}\right)\right.\nonumber\\
    & \left. m_q^4 \frac{1-3x+3x^2-(1-x)x}{x^3 (1-x)^3}\frac{\partial}{\partial m^2}\Lambda_3 \left(p^2,m_{\chi_j^0}^2,m_h^2,\frac{m_q^2}{x(1-x)}\right)\right]
\end{align}

with the $X_n,Y_n\,,n=1,2,3$ functions defined by:
\begin{align}
    & \int \frac{d^4 k}{(2\pi)^4}\frac{1}{\left(k^2-\frac{m_q^2}{x(1-x)}\right)^n \left[(p+k)^2-M^2\right](k^2-m^2)}=\frac{i}{(4\pi)^2}X_n\left(p^2,M^2,m^2,\frac{m_q^2}{x(1-x)}\right)\nonumber\\
    & \int \frac{dk^4}{(2\pi)^4}\frac{k^\mu}{\left(k^2-\frac{m_q^2}{x(1-x)}\right)^n \left[(p+k)^2-M^2\right](k^2-m^2)}=\frac{i}{(4\pi)^2}p^\mu Y_n\left(p^2,M^2,m^2,\frac{m_q^2}{x(1-x)}\right)
\end{align}

\section{Loop functions for the 2HDM+$a$}
\label{sec:loop_CP_odd}

In this appendix we summarize the main expressions, relevant for the computation of the SI scattering cross-section of the 2HDM+a, as given in \cite{Abe:2018emu}.
The coefficient associated to the triangle shaped topology is given by:
\begin{align}
    & C_q^{\rm triangle}=-\frac{m_\chi y_\chi^2}{(4\pi)^2}\left \{ \lambda_{\phi aa} \cos^2 \theta \, C_2(m_\chi^2,m_a^2,m_\chi^2) +\lambda_{\phi AA} \sin^2 \theta \, C_2(m_\chi^2,m_a^2,m_\chi^2)\right.\nonumber\\
    & \left. + \frac{\lambda_{\phi aA} \sin 2 \theta}{m_A^2-m_a^2} \left[B_1 (m_\chi^2,m_A^2,m_\chi^2)-B_1 (m_\chi^2,m_a^2,m_\chi^2)\right]\right \},
\end{align}
with $\lambda_{\phi XY},\,\,X, Y=a,A$ being the trilinear coupling between on CP-even and two CP-odd scalars. The function $C_2$ has been given in the previous appendix while the function $B_1$ is defined by the following integral:
\begin{equation}
    \int \frac{d^4 k}{(2\pi)^4}\frac{k^\mu}{(k^2-m^2)\left[(k+p)^2-M^2\right]}=\frac{i}{(4\pi)^2}p^\mu B_1 (p^2,M^2,m^2)
\end{equation}
The functions associated to the box diagrams are given by:

\begin{align}
    & C_q^{\rm box}=-\frac{m_\chi y_\chi^2}{(4\pi)^2}{\left(\frac{m_q}{v_h}\right)}^2\left \{\frac{\lambda_{Aqq}^2 \sin^2 \theta}{m_A^2}\left[G(m_\chi^2,0,M_A^2)-G(m_\chi^2,M_A^2,0)\right]\right. \nonumber\\
    & \left. +\frac{\lambda_{aqq}^2 \cos^2 \theta}{M_a^2}\left[G(m_\chi^2,0,M_a^2)-G(m_\chi^2,M_a^2,0)\right]+\frac{\lambda_{aqq}\lambda_{Aqq} \sin \theta \cos \theta}{M_A^2-M_a^2}\left[G(m_\chi^2,0,M_A^2)-G(m_\chi^2,M_a^2,0)\right] \right \},\nonumber\\
    & C_q^{(1) \rm box}=-\frac{8 y_\chi^2}{(4\pi)^2}{\left(\frac{m_q}{v_h}\right)}^2 \left \{\frac{\lambda_{Aqq}^2 \sin^2 \theta}{m_A^2}\left[X_{001}(p^2,m_\chi^2,0,M_A^2)-X_{001}(p^2,m_\chi^2,m_A^2,0)\right]\right. \nonumber\\
    & \left. +\frac{\lambda_{aqq}^2 \cos^2 \theta}{m_a^2}\left[X_{001}(m_\chi^2,m_\chi^2,0,M_a^2)-X_{001}(m_\chi^2,m_a^2,0)\right]+ \frac{\lambda_{aqq}\lambda_{Aqq} \sin \theta \cos \theta}{m_A^2-m_a^2}\left[X_{001}(m_\chi^2,m_\chi^2,0,m_A^2)-\right.\right. \nonumber\\
    &\left. \left. X_{001}(m_\chi^2,m_\chi^2,m_a^2,0)\right] \right \},\nonumber\\
    & C_q^{(2) \rm box}=-\frac{8 y_\chi^2}{(4\pi)^2}{\left(\frac{m_q}{v_h}\right)}^2\left \{\frac{\lambda_{Aqq}^2 \sin^2 \theta}{M_A^2}\left[X_{111}(m_\chi^2,m_\chi^2,0,m_A^2)-X_{111}(m_\chi^2,m_\chi^2,m_A^2,0)\right]\right. \nonumber\\
    & \left. +\frac{\lambda_{aqq}^2 \cos^2 \theta}{m_a^2}\left[X_{111}(m_\chi^2,m_\chi^2,0,m_a^2)-X_{111}(p^2,m_\chi^2,m_a^2,0)\right]+ \frac{\lambda_{aff}\lambda_{Aqq} \sin \theta \cos \theta}{m_A^2-m_a^2}\left[X_{111}(m_\chi^2,m_\chi^2,0,m_A^2)-\right.\right. \nonumber\\
    &\left. \left. X_{111}(m_\chi^2,m_\chi^2,m_a^2,0)\right] \right \},\nonumber\\
    & C_G^{\rm box}=\sum_{q=c,b,t}\frac{-m_\chi y_\chi^2}{432\pi^2}{\left(\frac{m_q}{v_h}\right)}^2\left[\lambda_{aqq}^2 \cos^2 \theta \frac{\partial F(m_a^2)}{\partial m_a^2}+ \lambda_{Aqq}^2 \sin^2 \theta \frac{\partial F(m_A^2)}{\partial m_A^2} +\lambda_{Aqq}\lambda_{aqq} \sin 2\theta \frac{\left[F(m_A^2)-F(m_a^2)\right]}{m_A^2-m_a^2}\right],
\end{align}
Where:
\begin{align}
    & G(m_\chi^2,m_1^2,m_2^2)=6 X_{001}(m_\chi^2,m_\chi^2,m_1^2,m_2^2)+m_\chi^2 X_{111}(m_\chi^2,m_\chi^2,m_1^2,m_2^2)\nonumber\\
    & \int \frac{d^4 k}{(2\pi)^4}\frac{k^\mu k^\nu k^\alpha}{\left[(k+p)^2-M^2\right](k^2-m_1^2)^2 (k-m_2^2)}=\frac{i}{(4\pi)^2}\left[\left(g^{\mu \nu}p^\alpha+g^{\nu \alpha}p^\mu+g^{\alpha \mu}p^\nu\right)X_{001}(p^2,M^2,m_1^2,m_2^2)\right.\nonumber\\
    &\left. +p^\mu p^\nu p^\alpha X_{111}(p^2,M^2,m_1^2,m_2^2)\right]
\end{align}
and:
\begin{align}
    & F(m^2)=\int_0^1 dx \left \{ 3 Y_1\left(m_\chi^2,m_\chi^2,m^2,\frac{m_q^2}{x(1-x)}\right)-m_q^2 \frac{2+5x-5x^2}{x^2(1-x)^2}Y_2\left(m_\chi^2,m_\chi^2,m^2,\frac{m_q^2}{x(1-x)}\right)\right. \nonumber\\
    & \left. -2 m_q^4 \frac{1-2x+2x^2}{x^3(1-x)^3}Y_3\left(m_\chi^2,m_\chi^2,m_a^2,\frac{m_q^2}{x(1-x)}\right)\right \}
\end{align}
where the $Y_{i=1,2,3}$ are the same given in the previous appendix.

\bibliographystyle{utphys}
\bibliography{biblio}
\end{document}